\documentclass[twocolumn,showpacs,preprintnumbers,amsmath,amssymb]{revtex4}

\usepackage{graphicx}
\usepackage{dcolumn}
\usepackage{bm}

\begin{document}
\title{Bose-Einstein correlations of $\boldsymbol{\pi^-\pi^-}$ pairs in central Pb+Pb collisions at CERN SPS energies}

\author{C.~Alt$^9$} 
\author{T.~Anticic$^{23}$}  
\author{B.~Baatar$^8$} 
\author{D.~Barna$^4$} 
\author{J.~Bartke$^6$} 
\author{L.~Betev$^{10}$} 
\author{H.~Bia{\l}\-kowska$^{20}$} 
\author{C.~Blume$^9$} 
\author{B.~Boimska$^{20}$} 
\author{M.~Botje$^1$} 
% ...

%
%

\affiliation{$^1$NIKHEF, Amsterdam, Netherlands }
% ...
\affiliation{$^4$KFKI Research Institute for Particle and Nuclear Physics, Budapest, Hungary}
% ...
\affiliation{$^6$Institute of Nuclear Physics, Cracow, Poland}
% ...
\affiliation{$^8$Joint Institute for Nuclear Research, Dubna, Russia}
\affiliation{$^9$Fachbereich Physik der Universit\"{a}t, Frankfurt, Germany}
\affiliation{$^{10}$CERN, Geneva, Switzerland}
% ...
\affiliation{$^{20}$Institute for Nuclear Studies, Warsaw, Poland}
% ...
\affiliation{$^{23}$Rudjer Boskovic Institute, Zagreb, Croatia}

\begin{abstract}
Measurements of Bose--Einstein correlations of $\pi^-\pi^-$ pairs in central
Pb+Pb collisions were performed with the NA49 detector at the CERN SPS for
beam energies of 20$A$, 30$A$, 40$A$, 80$A$, and 158$A$~GeV. Correlation
functions were measured in the longitudinally co-moving ``out-side-long'' reference frame
as a function of rapidity and transverse momentum in the forward hemisphere
of the reaction. Radius and correlation strength parameters were obtained from fits
of a Gaussian parametrization. The results show a decrease of the radius
parameters with increasing transverse momentum characteristic of strong
radial flow in the pion source. No striking dependence on pion-pair rapidity
or beam energy is observed. Static and dynamic properties of the pion source
are obtained from simultaneous fits with a blast-wave model to radius 
parameters and midrapidity transverse momentum spectra. Predictions
of hydrodynamic and microscopic models of Pb+Pb collisions are discussed.

\centerline{(Septemer 26, 2007)}
\end{abstract}


\maketitle

\section{Introduction}
\label{intro}

In central collisions of ultra-relativistic heavy ions an extended volume 
of high energy density is created. Several observations at top SPS  \cite{Heinz:2000bk} 
and RHIC \cite{Adcox:2004mh,Back:2004je,Adams:2005dq,Arsene:2004fa}  
energies indicate that the initial energy density is large enough to overcome QCD confinement 
and to form a phase of deconfined quarks and gluons during the early stages of the
collisions.

This strongly compressed interacting matter expands into the surrounding vacuum, causing
the temperature and energy density to drop to the critical values and forcing the system to form hadrons 
at the phase boundary. During the subsequent stage the hadrons may continue to scatter inelastically
and elastically until all interactions cease.

The excitation functions of several observables, e.g. the energy dependence
of the pion yield per wounded nucleon and the strange particle per pion ratio
indicate that deconfinement indeed appears already in central Pb+Pb collisions 
at lower SPS energies \cite{Gazdzicki:2004ef}. The transient existence of a deconfined phase
in the early stage of the reaction may be still reflected in such properties as e.g. the
lifetime, final size, and collective flow of the freeze-out stage of the reaction \cite{Pratt:1986cc,Bertsch:1989vn}.

The study of Bose-Einstein (BE) correlations has been shown to contribute unique information
on the static and dynamic properties of the system at thermal freeze-out
\cite{Wiedemann:1999qn,Lisa:2005dd}. Traditionally a multi-dimensional Gaussian function,
parametrized by radius parameters, is fitted to the BE correlation function.
Quantitative parameters of the pion source can be derived from a
simultaneous analysis of single-particle transverse momentum spectra and two-particle 
BE radius parameters  using hydrodynamically inspired models describing the evolution of the system.
Detailed measurements of the rapidity and transverse momentum
dependence of BE correlations for $\pi^-\pi^-$ pairs in the full SPS energy range from 20$A$ to 158$A$
GeV beam energy and their interpretation in terms of a rapidly expanding fireball
are the subject of this paper. An alternative analysis scheme of BE correlations has been
developed \cite{Brown:1997ku,Brown:1997sn}, which tries to deduce the
properties of the pion source directly from the BE correlations by an integral inversion technique.
An analysis using this method has been applied to RHIC data~\cite{Adler:2006as} and is currently in progress for the SPS
data of NA49.

An earlier analysis of  $\pi^{-}$$\pi^{-}$ BE correlations at 158{\it A} GeV beam energy with
the NA49 spectrometer \cite{Appelshauser:1997rr} pioneered the simultaneous analysis of
BE correlations and inclusive transverse momentum spectra. The study
found a mean velocity of the collective transverse expansion 
of $<\beta> = 0.55 \pm 0.12$ and a freeze-out temperature of $T=120 \pm 12$ MeV. Furthermore a strong longitudinal 
expansion and a finite duration of particle emission of 3 to 4 fm/c were observed. The results presented
in this paper basically agree with these findings. One of the major differences of the two analyses lies in the parametrization
of the correlation function: in the present work the momentum difference is decomposed into Cartesian components
$q_{side}$, $q_{out}$, and $q_{long}$ as suggested by Podgoretsky \cite{Podgoretsky:1982xu},
Pratt \cite{Pratt:1986cc}, and Bertsch \cite{Bertsch:1988db}; in the previous publication the parametrization 
developed by Yano-Koonin \cite{Yano:1978gk} and Podgoretsky \cite{Podgoretsky:1982xu} was applied.   
In addition, in this article also measurements at lower SPS energies will be presented.

Results of an analysis of K$^+$K$^+$ and K$^-$K$^-$ BE correlations at 158{\it A} GeV beam energy
from NA49 were reported in \cite{Afanasiev:2002fv}. These are less differential due to the much smaller number 
of kaon pairs. However, taking account of the radial flow and comparing the extracted source parameters 
at equivalent values of the transverse masses, consistency is found between results from pion and kaon analyses.

The NA49 detector allows to study also correlations of heavier particles. Two-proton correlations
at 158{\it A} GeV beam energy were analysed in \cite{Appelshauser:1999in}. Here the contributions
from strong interactions, Coulomb repulsion and Fermi-Dirac quantum statistics are of similar magnitude and make it
impossible to extract source parameters with the help of a simple parametrization. The final result of that study, 
i.e. the extracted averaged source radius, is compatible with that obtained from $\pi^{-}\pi^{-}$ and $\text{K}^{\pm}\text{K}^{\pm}$ BE correlations.

Numerous other $\pi\pi$ BE correlation analyses near midrapidity were reported in the literature.
Measurements at the AGS can be found in \cite{Lisa:2000hw}. Results from the SPS were published by
experiments NA44 \cite{Bearden:1998aq}, WA97 \cite{Antinori:2001yi}, WA98 \cite{Rosselet:2002ru}, and
NA45 \cite{Adamova:2002wi}. Measurements at RHIC have been reported by STAR \cite{Adler:2001zd,Adams:2004yc},
PHENIX \cite{Adler:2004rq} and PHOBOS \cite{Back:2004ug}.
Results show good consistency in general and exhibit remarkably little change over this very
large collision energy range. At all energies, hydrodynamic models were found to be able to consistently describe 
many aspects of the reactions, including the transverse momentum dependence of BE~correlations
\cite{Heinz:2001xi}. 
The exception remains the over-prediction of the experimentally determined
pion source radii by several fermi. This could be due to a too simplistic treatment of
the freeze-out stage \cite{Borysova:2005ng}
and/or a neglect of possible refractive effects in the fireball within
the traditional interpretation of BE~correlations \cite{Wong:2004gm,Kapusta:2004ju,Cramer:2004ih}.

This paper is organized as follows. In section 2 the NA49 experimental setup is described followed by
a brief introduction to BE correlations in section 3.
Selection criteria for events, tracks, and pairs are presented in section 4. Corrections, systematic
uncertainties, and the fitting procedure are described in section 5. Results are presented in section 6.  The paper is closed
by a discussion in section 7 and conclusions in section 8.

\section{Experimental setup}

\label{NA49 Experiment}

\begin{figure*}
\centering
\resizebox{0.9\textwidth}{!}{
\includegraphics{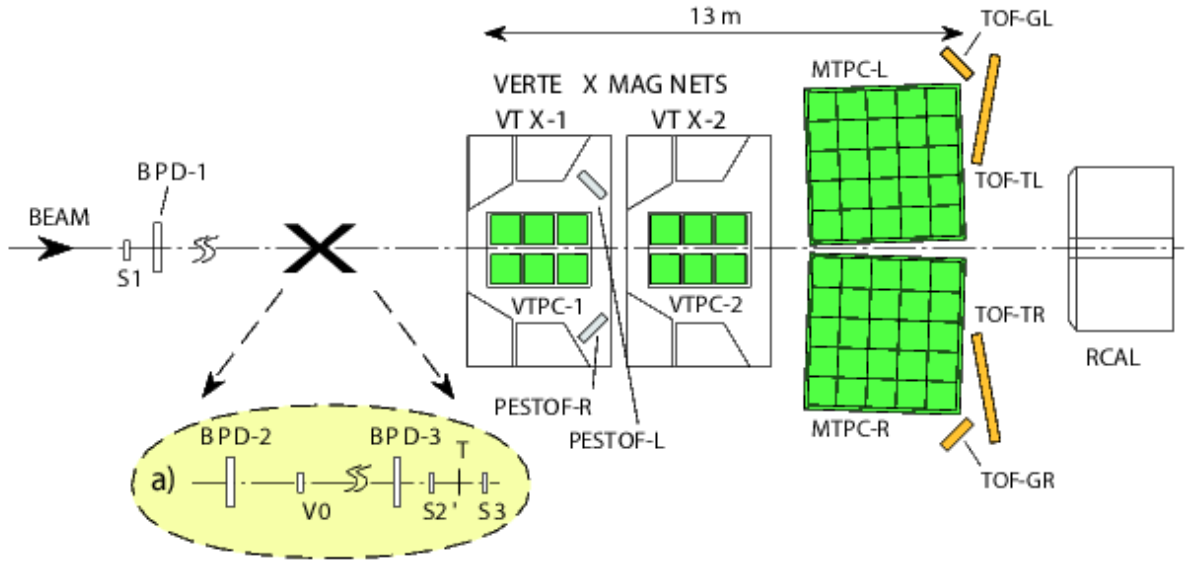}}
\caption{(Color online)~The NA49 experiment: beam position detectors BPD, target foil T, time of flight detectors TOF, time projection chambers
TPC, and calorimeters CAL. Details can be found in \cite{Afanasev:1999iu}.}
\label{na49layout}       
\end{figure*}

The layout of the essential components of the NA49 detector is displayed in Fig.~\ref{na49layout}.

Trajectories of incident beam particles are measured individually by beam position detectors
BPD 1-3. Collisions occur in a thin lead foil T of 224 mg/cm$^2$ thickness. The main tracking devices 
for charged particles are four large {\bf{T}}ime {\bf{P}}rojection {\bf{C}}hambers (TPC), 
labeled VTPC1, VTPC2, MTPCL, and MTPCR in Fig.~\ref{na49layout}. 
Two of them, VTPC1 and VTPC2, are mounted in precisely mapped magnetic fields provided by two superconducting magnets with total
bending power of up to 9 Tm. The TPCs provide large acceptance, efficient reconstruction 
and precise tracking for charged particles. The momentum of a charged particle can be determined
from its characteristic deflection in the magnetic field; with the NA49 setup a resolution 
of $\Delta p / p^2 \approx (0.3-7) \cdot 10^{-4}$~$\rm{(GeV/c)^{-1}}$ is achieved.
A veto calorimeter VCAL behind a suitably adjusted collimator COLL measures the energy of the projectile
spectators. The event trigger consists of a valid beam track signal in coincidence with an energy deposition
in the veto calorimeter below a preset threshold. Further details of the detector and trigger system
can be found in \cite{Afanasev:1999iu}.
The detection efficiency depends on particle momentum and varies from about 90\% at midrapidity to near 100\% in
the forward direction. \\
\begin{figure}[h!]
\centering
\resizebox{0.45\textwidth}{!}{
\includegraphics{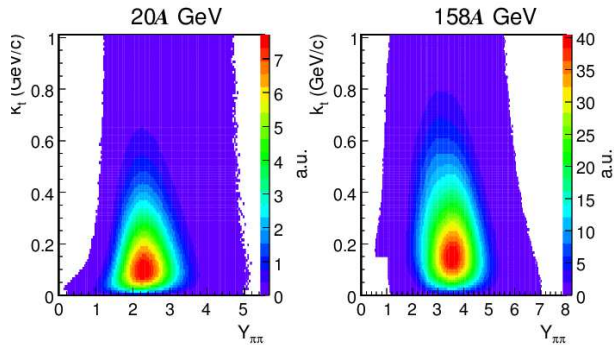}}
\caption{(Color online)~Raw yields of particle pairs at 20$A$ and 158$A$ GeV beam energy as a function of ${Y_{\pi\pi}}$ and ${k_{t}}$.}
\label{acceptance}
\end{figure}
The strength of the magnetic field was scaled according to beam energy, i.e. at 20{\it A} GeV 
it was set to $\approx$1/8 
of the maximum setting at 158{\it A} GeV. In this way, the geometric acceptance in the center-of-mass
system is kept similar at different beam energies.  
Fig.~\ref{acceptance} displays the raw yields of particle pairs at 20{\it A} GeV and at 158{\it A} GeV 
beam energy.

The forward hemisphere in the center of mass of the reaction is well covered allowing differential studies in 
pion-pair rapidity 
\begin{equation}
 {Y_{\pi\pi}}={{1\over2}ln{{E_1 + E_2 + p_{z,1} + p_{z,2}}\over{E_1+E_2-p_{z,1}-p_{z,2}}}}
\end{equation}
calculated in the reaction c.m. system and averaged transverse momentum 
\begin{equation}
{k_{t}}={{1\over2}|\bold{p_{t,1}}+\bold{p_{t,2}}|}, 
\end{equation}
where $E_i$ and $\bold{p_i}$ represent
energies and momenta of the pions.
Track topology and momentum resolution change with transverse momentum, rapidity, and different settings 
of the magnetic field.  The impact of the finite momentum resolution on the correlation function 
is discussed in detail in section~\ref{sec:momentum-resolution}.

Along with their trajectory, the NA49 TPCs measure the specific energy loss of charged particles traversing the detector gas.
Combined with the reconstructed momentum this information allows to 
identify the particle type. This selection is not applied in the present analysis since the procedure works only 
in a limited momentum range.
The impact of the missing particle identification on the correlation function is discussed
in section~\ref{sec:missing-pid}.

The energy deposition in the veto calorimeter corresponds to the total energy of the projectile spectators and is
closely related to the centrality of the collision.
At lower beam energies the threshold of the online trigger was set to accept the 7.2\%  most central events.
The dataset at 158{\it A} GeV was taken with a 10\% online trigger setting, but to ease the comparison the
event selection was restricted offline to the 7.2\% most central events. The average number of wounded nucleons for
the 7.2\% most central collisions was estimated by Glauber model calculations and amounts to 
$\langle N_W \rangle$ = 349 with a dispersion of 28.

In the course of the CERN SPS energy scan programme the NA49 experiment used
$^{208}\rm{Pb}$ beams with energies of  20{\it A}, 30{\it A}, 40{\it A}, 
80{\it A}, and 158{\it A} GeV. 
A summary of the central Pb+Pb data sets used in this analysis 
is given in Table~\ref{table:escan}. 

\begin{table}[h]
\caption{ Data sets used in this analysis.}
\label{table:escan}

\centering
\begin{tabular}{c|c|c|c|c}
\parbox[c]{0.95cm}   {\scriptsize{Beam Energy \newline (GeV)} }   
& \parbox[c]{0.80cm} {\scriptsize{$\sqrt{s_{NN}}$ \newline (GeV)} }
& \parbox[c]{1.0cm}  {\scriptsize{Year of \newline data taking} }
& \parbox[c]{1.6cm}  {\scriptsize{Number of \newline analyzed events} }  
& \parbox[c]{1.2cm}  {\scriptsize{Mag. field \newline Polarity} } \\ \\
\hline

20  & 6.3  & 2002 & 360k & +\\
30  & 7.6  & 2002 & 420k & +\\
40  & 8.7  & 2001 & 217k & -\\
40  & 8.7  & 2000 & 360k & +\\
80  & 12.5 & 2001 & 296k & +\\
158 & 17.3 & 1996 & 386k & +\\
158 & 17.3 & 2000 & 502k & - \\

\end{tabular}   
\end{table}

\section{Bose-Einstein correlation function}
\label{chapter:intro_hbt}
BE correlations are a consequence of the requirement that the quantum mechanical
wave function of two identical bosons must be symmetric under particle exchange. This results in  an increased
probability $P_{1,2}$ of emitting pion pairs at small momentum difference $q=p_{1}-p_{2}$ relative to 
the product of probabilities $P_{1}\cdot P_{2}$ to emit single pions of momentum $p_{i}$.
The enhancement is usually studied in terms of the correlation function: 

\begin{equation}
  {C(q)} = {{P_{1,2}} \over {P_{1}\cdot P_{2}}} = {{S(q)} \over {B(q)}} .
\label{equation:C_exp}
\end{equation}

Experimentally the correlation function is constructed as the ratio of the pair momentum difference distribution 
$S(q)$ and a mixed-event pair distribution $B(q)$. This reference distribution is constructed by
forming pairs of particles from different events.
Except for genuine particle correlations, $B(q)$ thus includes the features of 
particle detection with the NA49 apparatus as well as the characteristics of inclusive
pion production from the source. The track selection criteria, described in the next section, are applied 
in the construction of the reference in the same way as in the construction of the signal distribution. 

The distribution of spatial distances of the emission points of the two pions is related 
to the distribution of their momentum difference. In fact, the spatial extension of the pion
emitting source is reflected in the width of the BE enhancement in the $\bold{q}$-dependent correlation function.
In this paper, the inverse of the width of the enhancement is labelled the BE radius. Elsewhere it is sometimes also referred to 
as HBT radius, in reference to the pioneering work of Hanbury Brown and Twiss \cite{HanburyBrown:1954wr}.

Since momenta and emission points of particles from the source are also correlated due to radial and longitudinal
collective flow, the measured radii correspond to lengths of homogeneity \cite{Makhlin:1987gm}
rather than to the actual extension of the source. This feature allows to extract dynamic parameters
of the source, if the correlation function is analyzed as a function of the kinematic variables
${Y_{\pi\pi}}$ and ${k_{t}}$.

The momentum difference is decomposed into 3 independent components as suggested by Podgoretsky \cite{Podgoretsky:1982xu}, 
Pratt \cite{Pratt:1986cc}, and  Bertsch \cite{Bertsch:1989vn}. The first,
${q_{long}}={p_{z,1}-p_{z,2}}$, is the momentum difference along the direction of the beam,
measured in the longitudinal co-moving system,
i.e. in the frame of reference where $p_{z,1}=-p_{z,2}$. The two other components are defined 
in the plane transverse to the beam, with
$\bold{q_{out}}$ parallel to the pair transverse momentum vector 
$\bold{k_{t}}={{1\over2} ( \bold{p_{t,1}} + \bold{p_{t,2} ) }}$ and $\bold{q_{side}}$ 
perpendicular to $\bold{q_{out}}$ and $\bold{q_{long}}$.

The two-particle BE correlation  function $C_{BE}$ is in general well approximated by a 
three dimensional Gauss function
\begin{eqnarray}
C_{BE}(\bold{q})=1+\lambda\cdot exp(&-&  q_{out}^2 R_{out}^2 - q_{side}^2 R_{side}^2 \nonumber \\ 
                                    &-&  q_{long}^2 R_{long}^2 \nonumber \\ 
                                    &-& 2q_{out}q_{long}R_{outlong}^2). 
\label{eq:C_BE}
\end{eqnarray}
A fit to the measured correlation function yields results for the variances
$R_{out}$, $R_{side}$, $R_{long}$, and $R_{outlong}$, which
can then be compared to results from model calculations.
For several reasons the measured correlation functions have amplitudes 
smaller than the theoretically expected value of two for vanishing momentum difference. 
The known sources of the suppression, such as contamination by misidentified particles
or by weak decay products, are taken into account by introducing a purity factor $p$, described in 
section~\ref{sec:missing-pid}. To allow for additional sources of suppression
a coherence parameter $\lambda$ is included in the fitting procedure.

\section{Event, track, and pair selection}
\subsection{Event cuts}
Events were accepted if they had a successfully reconstructed beam particle and interaction vertex and an energy 
deposition in the veto calorimeter within the appropriate range. The position of the main vertex was
determined from the set of reconstructed tracks. Its position had to be close to that of the beam track 
in the transverse plane and close to the known position of the target foil in the longitudinal direction.
The deviations of the main vertex coordinates from the expected positions
show Gaussian distributions in x (bending plane), y (vertical), and z (longitudinal) directions. 
Events are accepted if their reconstructed interaction points are not farther away than three 
standard deviations from the respective mean values.

\subsection{Track cuts}
In this analysis, only negatively charged particles 
satisfying a set of quality criteria on measured track length and distance of closest approach
to the event vertex were used.

Usually tracks with less than a minimum number of measured points are excluded 
from physics analyses of NA49 data. 
However, it was found that in the VTPC1 detector tracking quality depends on the occupancy of the TPCs
which may vary even for fixed multiplicity in an event due to
spiraling electrons generating large numbers of clusters in a TPC. These 
kinds of fluctuations of the performance of the detector cause a subtle dependence 
of the correlation function on the
minimum number of required measured points. In order to avoid this problem
the track quality was ensured by requiring a minimum number of 20 potential points 
simultaneously in one of the VTPCs and one of the MTPCs out of a possible 
maximum of 72 and 90 points respectively. The number of potential points is
derived from a detailed simulation and reconstruction of particle tracks in the detector.
It corresponds to the number of charge clusters detectable in the TPC under ideal conditions. 
Varying this cut over a reasonable interval had no influence on the finally extracted correlations.

To minimize contributions from products of weak decays and from secondary interactions with the 
detector material only those tracks were accepted that appeared to originate from
the main vertex. Extrapolations of tracks to the nominal position of the target
were required to deviate not more than 5.0 cm in x-direction and less than 3.0 cm 
in y-direction from the reconstructed position of the main vertex.

\subsection{Pair cuts}
\subsubsection{Two track resolution cut}
\label{chap:Two track resolution cut}
Since this study focuses on particle correlations at low momentum difference the two-track
resolution capabilities of the detectors were investigated carefully.
Tracks of particles with small momentum difference travel closely together in the TPCs. If two tracks approach each other
the clusters may start to overlap.
In this case their charge deposition distributions are not always resolved properly and
the ionization points are not sufficiently well reconstructed. In extreme cases measured points might be assigned to the wrong
track changing the measured values of momentum differences significantly.  
Particle pairs possibly affected by the limited two-track resolution are therefore excluded from the analysis.

\begin{figure}[h]
\centering
\resizebox{0.45\textwidth}{!}{
\includegraphics{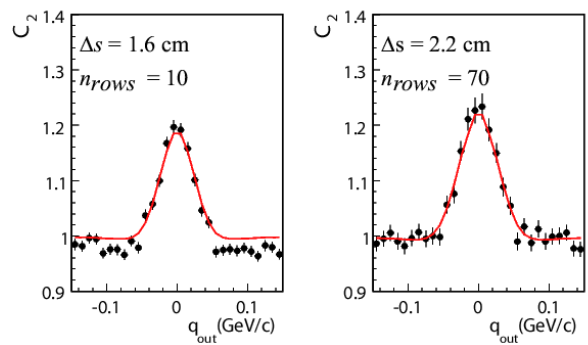}}

\caption{(Color online)~Projections of the correlation function on the $q_{out}$ component 
at 158$A$~GeV beam energy at midrapidity in the transverse momentum bin  $0.3<k_{t}<0.4$~GeV/c. 
Left hand side: requiring a distance of closest approach DCA of 1.6~cm at 10 consecutive padrows, 
at the right hand side: a DCA of 2.2~cm was required at 70 padrows.}
\label{fig:dcacut_oldnew}
\end{figure}

In preliminary analyses of NA49 data a simple anti-merging cut was applied, based on 
an ``averaged two-track distance''. The distance of two 
tracks in the transverse plane was calculated at two or more fixed planes, e.g. at entrance and exit of a TPC, and the average of 
these values was defined as the two-track distance. Only pairs were accepted, which had 
two-track distances larger than a certain minimum value.
A detailed study revealed, that this kind of cut is not able to remove all corrupted pairs, resulting in a drop below unity in the
projection of the  $q_{out}$ component of the correlation function visible on the left hand side of Fig.~\ref{fig:dcacut_oldnew}.
Therefore a more involved procedure was developed which will be described next.

\begin{figure}[h]
\centering
\resizebox{0.45\textwidth}{!}{
\includegraphics{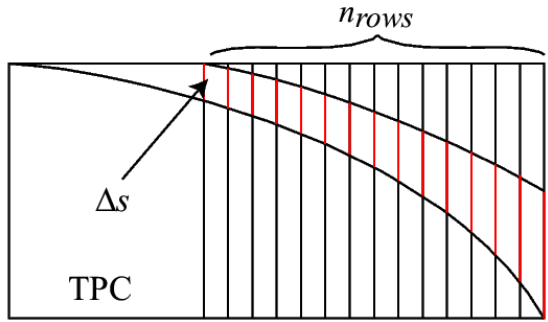}}

\caption{(Color online)~Schematic description of the anti-merging cut: two tracks
        have to be separated by more than $\Delta s = 2.2$ cm at 
        more than $n_{rows} = 50$ consecutive padrows.}
\label{fig:dcacut}
\end{figure}

\begin{figure}[h]
\centering
\resizebox{0.45\textwidth}{!}{
\includegraphics{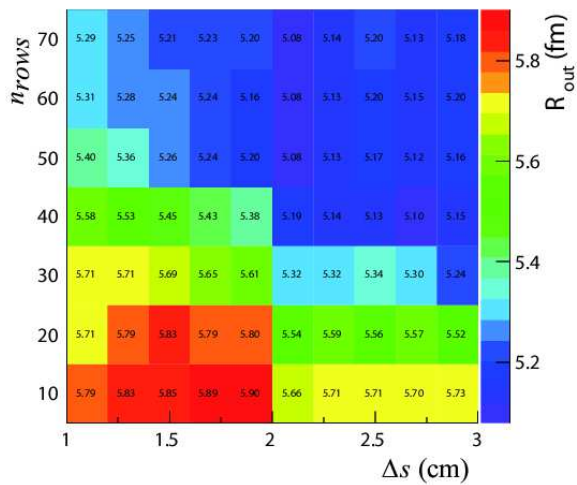}}
\caption{(Color online)~Dependence of the parameter $R_{out}$ on the 
        minimum separation $\Delta s$ and on the required number of rows, at midrapidity at $0.3<k_{t}<0.4$~GeV/c for the 158$A$~GeV beam energy sample.}
\label{fig:routvsdcacut}
\end{figure}

First each particle trajectory is extrapolated through the TPCs and the transverse positions at 
the nominal longitudinal positions of the individual padrows are stored. 
Thus, for every pair the distance at each padrow is readily calculable. The trajectories are required
to be separated by more than $\Delta s = 2.2$~cm at more than 50 consecutive padrows, starting 
at the end of each TPC (see Fig.~\ref{fig:dcacut}). 
If one of the tracks of a pair has less than 50 potential points, the criterion must be fulfilled at all
padrows where both tracks have a potential point.
Starting from VTPC1, this criterion is tested 
for each TPC in which both tracks have at least 20 potential points. If the requirement is not fulfilled in all TPCs, 
the pair is rejected. The impact of this cut on the finally extracted $R_{out}$ parameter is shown in 
Fig.~\ref{fig:routvsdcacut}. Varying this selection criterion can change the finally extracted parameters 
by up to 5\%.

\subsubsection{Split track cut}
\begin{figure}[!h]
\centering
\resizebox{0.45\textwidth}{!}{
\includegraphics{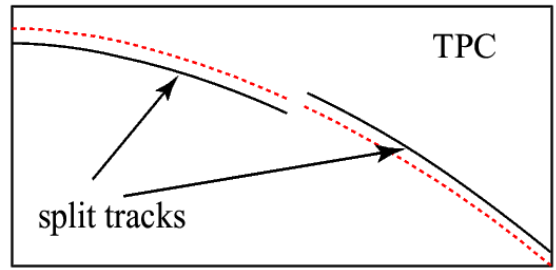}}
\caption{(Color online)~Schematic description of a split track: a single track (dashed line) is reconstructed as two different tracks (solid lines). }
\label{fig:Split_tracks}
\end{figure}

If not all charge clusters induced by a particle traversing a TPC are assigned to the track during the reconstruction 
of the track, it is possible that the remaining clusters are reconstructed as a separate track. In this case of ``split tracks''(see Fig.~\ref{fig:Split_tracks}) 
additional pairs of low momentum difference are artificially created when constructing the signal distribution. 
They alter the correlation function at low momentum difference significantly.

The procedure to identify and remove pairs composed of the two segments of a split track is based on the following considerations:
since both segments originate from the same particle they have approximately the same number of potential points $n_{pot}$. 
The sum of the measured points $n_{meas}(i)$ of both track segments  is equal to or smaller than the number of 
the potential points. Therefore, a corrupted pair most probably has:

\begin{eqnarray}
\frac{n_{meas}(1)}{n_{pot}(2)} + \frac{n_{meas}(2)}{n_{pot}(2)} \leq 1.0.
\label{eqn:split}
\end{eqnarray}

Due to possible fluctuations in the determination of the potential and measured number of points 
the cut value is increased from 1.0 to  1.1, to ensure the removal of all split-track pairs.
The criterion is checked for each TPC separately.

\subsection{Examples of correlation functions}

Correlation functions are constructed according to Eq.~\ref{equation:C_exp} as a three dimensional histogram with
10 MeV/c wide bins. The number of mixed events is chosen such that the background
contains about eight times more entries than the signal, making the contribution
from the background to the statistical error in the correlation function negligible. 
Due to the large number of real pairs, the relative statistical error attributed to numerical
values of parameters is small.
The whole sample of particle pairs is subdivided into kinematic bins for a differential analysis. 
A bin width of 0.5 units in pair rapidity ${Y_{\pi\pi}}$ is chosen, resulting in up to five bins in this direction,
covering the range from central rapidity up to beam rapidity at the highest beam energy.
In the transverse direction the ${k_{t}}$-value of the pair is used for the binning; 
five bins are defined with the boundary values 0.0, 0.1, 0.2, 0.3, 0.4, and 0.6 GeV/c.
With decreasing beam energy and increasing transverse momentum and pair rapidity the number 
of pair entries in the signal distribution
drops rapidly preventing the fitting procedure from converging; hence fit results are not available
in every kinematic bin at each beam energy.

\begin{figure}[h]
\centering
\resizebox{0.45\textwidth}{!}{
\includegraphics{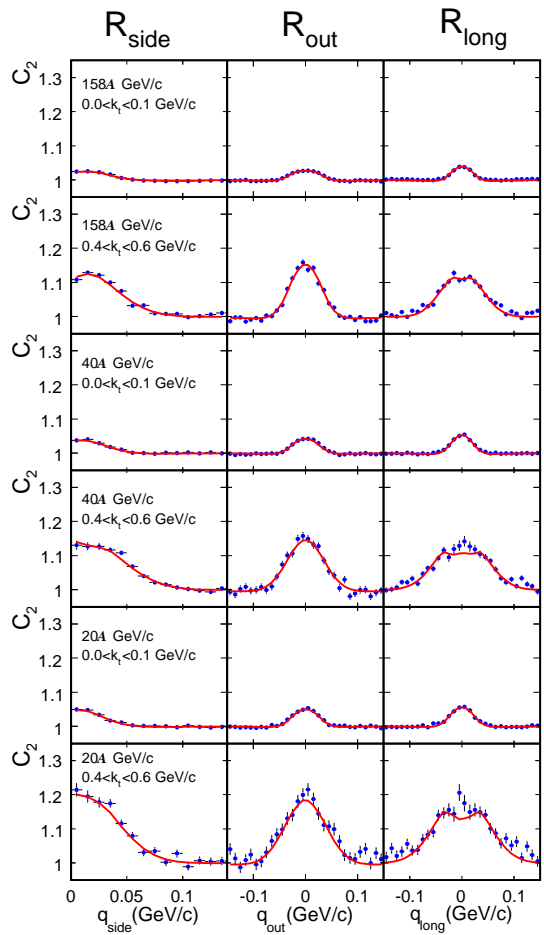}}
\caption{(Color online)~Projections of correlations functions at midrapidity at three different beam energies at 
low ($0.0<k_{t}<0.1$~GeV/c) and high ($0.4<k_{t}<0.6$~GeV/c) transverse momentum. The projected 
components have been integrated over 50 MeV/c. The curves show the projections of the 
fits of Eq.~\ref{eq:finalfitting} to the data. }
\label{fig:Projections}
\end{figure}

Fig.~\ref{fig:Projections} shows expamles of projections of the measured correlation
functions on the momentum difference components  $q_{side}$, $q_{out}$, $q_{long}$, 
with the integration range 0 - 50 MeV/c in both projected components.   
The examples were obtained for midrapidity, at 20{\it A}, 40{\it A}, and 158{\it A} GeV beam energy and
at low ($k_{t}<0.1$~GeV/c) and higher ($0.4<k_{t}<0.6$~GeV/c) averaged pair transverse momenta.
The solid lines represent the fits of Eq.~\ref{eq:finalfitting} to the 3-dimensional histograms, 
projected in the same way as the measured correlation functions.

Strong fluctuations of global event characteristics, e.g. of multiplicity or mean pair transverse momentum,
can distort the reference distributions. In this analysis no indications of this kind 
have been observed, hence all events are sampled equally when constructing the mixed pair
distributions.

\section{Corrections, fit procedure and systematic uncertainties}

The measured correlation function is a convolution of BE correlations with
several physics and detector related effects.

The finite momentum resolution of the NA49 detector distorts the correlation function.
The implications for the fitted parameters are discussed 
in section~\ref{sec:momentum-resolution}.  
The Bose-Einstein enhancement in the measured correlation function is diluted by
pairs consisting of at least one particle from a secondary interaction, from a weak decay,
or a negatively charged particle of a different identity (K$^-$, $\bar{\text{p}}$, e$^-$).
These effects are discussed in section~\ref{sec:missing-pid} and are corrected for by 
introducing a purity factor in the fitting procedure.

The two particle correlation function contains not only BE correlations, 
but also correlations due to the Coulomb repulsion between two negatively charged
particles. Since the aim is to extract correlation lengths due to the Bose-Einstein effect, 
a correction procedure is applied, which is described in section~\ref{sec:coulomb}. 

Further contributions to the correlation function, e.g. due to the strong interaction or
due to conservation laws, are assumed to be negligible in this analysis.
Technical aspects of the fitting procedure are discussed in section~\ref{sec:fitting}.

\subsection{Momentum resolution correction}
\label{sec:momentum-resolution}
The accuracy of the momentum determination is limited by the finite spatial resolution of the TPCs  
and by multiple Coulomb scattering of the particles with detector material.
The combined result is an uncertainty in 
the momentum determination which increases with increasing transverse momentum of the particle. 
For measurements at higher beam energies the decreasing contribution from multiple scattering leads
to a slight improvement of the momentum determination.

The finite momentum resolution causes a broadening of the correlation function and hence a reduction of 
the values of the fitted radius parameters. 
The size of the effect was determined from a detailed simulation in which 
the resolution of the momentum difference determination 
$\bold{ \delta q} (\delta q_{out}, \delta q_{side}, \delta q_{long})$ for each $\bold{q} (q_{out}, q_{side}, q_{long})$ 
bin was derived from a Monte Carlo study modeling the whole particle detection process and the reconstruction algorithm.

The simulation allows to compare the generated momentum difference of a particle pair $q^{gen}$ with 
the reconstructed value $q^{rec}$. The differences in each component, e.g. $q_{out}^{rec} - q_{out}^{gen}$, 
are filled into frequency distributions. 
The width of the frequency distribution defines the relative resolution in the respective component, e.g. $\delta q_{out}$. 
Since the momentum resolution depends on the particle momenta (track curvature in the TPCs) 
the procedure has to be performed
differentially for each $\bold{q} (q_{out}, q_{side}, q_{long})$ bin in each ($Y_{\pi\pi}$,$k_{t}$) interval 
and has to be redone for each beam energy sample.
In the worst case, i.e. at the lowest energy and for the $q_{out}$ component, 
the mean resolution $\langle \delta  q_{out} \rangle$ is still $\sim$5 MeV/c. 



The impact of the finite momentum resolution on the correlation function is estimated 
in the following way: two artificial correlation functions $C_{ideal}$ and $C_{smear}$ are generated with their ratio
equal to the ratio of the ``true'' correlation function ${C_{true}}$ to the measured one ${C_{meas}}$:

\begin{equation}
  {{C_{true}} = {{C_{meas}} \cdot {{C_{ideal}} \over {C_{smear}}}}}.                    
\label{eq:C_smear}
\end{equation}

Multiplying the measured correlation function with the ratio ${C_{ideal}}/{C_{smear}}$
allows to extract the BE parameters corrected for finite momentum resolution. 

$C_{ideal}=S_{ideal}/B_{ideal}$ represents the correlation function measured by an ideal detector.
It is derived from the BE-correlation function given by Eq.~\ref{eq:C_BE} assuming 
BE-parameters close to the fitted ones. Technically, the signal distribution is generated 
by weighting the measured mixed-pair entries according to Eq.~\ref{eq:C_BE}: 
$S_{ideal} = B_{meas} \cdot C_{BE}(q^{gen})$ and the background distribution is simply defined as
$B_{ideal} = B_{meas}$ yielding $C_{ideal}=S_{ideal}/B_{meas}$.

In the construction of the correlation function $C_{smear}$, the weight is calculated in the same way,
but the position $\bold{q_{smear}}$, where the entry is filled, is obtained from a smearing procedure 
reflecting the uncertainty in the momentum determination. This procedure results in
$C_{smear}(q_{smear})=S_{ideal}(q^{gen})/B_{meas}(q^{gen})$.
The smearing algorithm takes as input the momentum difference $q^{gen}$
of the pair and varies it slightly, using a random generator based on a Gaussian probability
distribution with variance $\delta q(\bold{q},k_{t},Y_{\pi\pi})$ and mean equal to $q^{gen}$.

\begin{figure}[!h]
\centering
\resizebox{0.45\textwidth}{!}{
\includegraphics{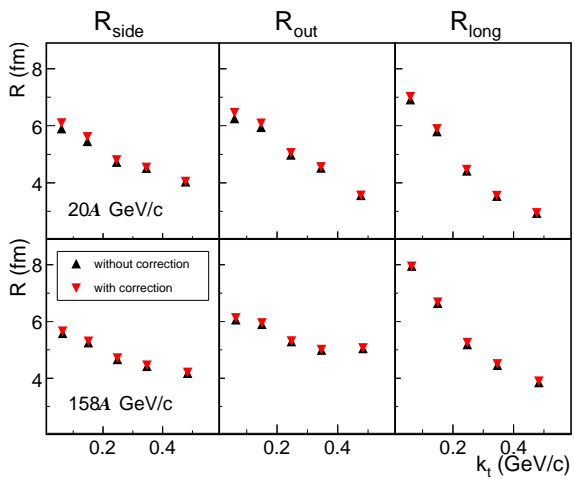}}
\caption{\label{momres} (Color online)~Fitted radii with (triangles up) and
without (triangles down) momentum semearing
at midrapidity at 20{\it A} (upper) and 158{\it A} GeV (lower row) respectively.}
\label{fig:fitswithmomres}
\end{figure}

The effect of momentum smearing on the extraction of the parameters is demonstrated in Fig.~\ref{fig:fitswithmomres}.
Radius parameters fitted with (triangles up) and without (triangles down) momentum smearing are
compared. Clearly there is little impact on the fit parameters for the NA49 experiment due to its
excellent momentum resolution. Hence a momentum resolution correction is not applied to the correlation 
functions used in the final fit.

\subsection{Missing particle identification and purity factor}
\label{sec:missing-pid}

No particle type selection was applied since the particle identification capabilities of the NA49 TPCs 
allow a sufficiently clean particle by particle identification only for 
laboratory momenta between roughly 5 and 50~GeV/c. 
This restriction prevents particle identification especially at lower transverse momenta at midrapidity. 
In other NA49 analyses this deficiency is compensated by using additional 
information from Time of Flight detectors. However, because of their limited acceptance this information is not employed in this 
analysis. Instead the correlation functions are determined for all negatively charged particle 
pairs h$^{-}$h$^{-}$ and corrected for contaminations.

\begin{figure}[h]
\centering
\resizebox{0.45\textwidth}{!}{
\includegraphics{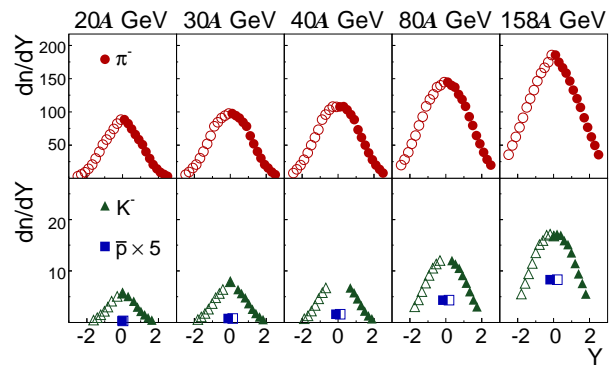}}
\caption{(Color online)~Rapidity distributions of negative pions and  kaons and the antiproton yield at midrapidity at SPS energies. 
Open symbols indicate reflected values \cite{Afanasiev:2002mx,Gazdzicki:2004ef,Alt:2006dk}. Data at 20{\it A} and 30{\it A} GeV are
preliminary results.}
\label{fig:rap_pi_ka}
\end{figure}

The fraction of pions in the sample of accepted particles depends on the energy of the collision and on the kinematic region.
Fig.~\ref{fig:rap_pi_ka} shows the rapidity distributions of pions and of the less abundant negatively charged kaons
and antiprotons at different beam energies. 
The pions represent about 90\% of the total number of negatively charged particles.

The dilution of the particle sample with non-pions reduces the correlation function. Since correlations between 
non-pions and any other particle, e.g. K$^{-}$K$^{-}$  or K$^{-}\pi^{-}$, can be neglected, 
the missing particle identification leads only to a reduction of the fraction of correlated pairs, 
which is taken into account by introducing the purity factor $p$:

\begin{equation}
  {C_{meas}}  =  {p \cdot C_{corr} + (1-p) }.
\label{equation:finalfit}
\end{equation}

$C_{meas}$ corresponds to the measured correlation function and $C_{corr}$ to the correlation function exclusively containing
correlated pairs in the signal distribution. The purity $p$ is defined as the ratio of the number of 
primary $\pi^-$ pairs to all negatively charged particle pairs in each bin of $Y_{\pi\pi}$ and $k_{t}$.  
 
\begin{figure}[h]
\centering
\resizebox{0.45\textwidth}{!}{
\includegraphics{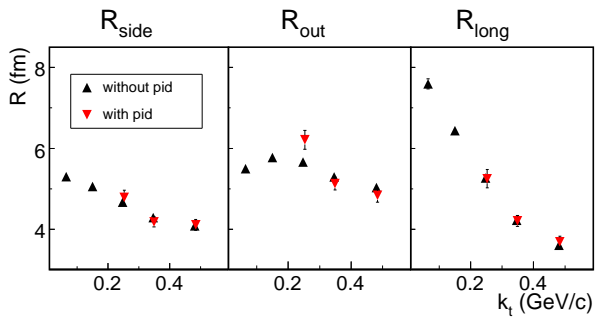}}
\caption{(Color online)~$k_{t}$ dependence of fit parameters at midrapidity at 158{\it A} GeV with and without particle identification.}
\label{fig:radii_withandwithout_dedx}
\end{figure} 

This procedure has been validated by comparing extracted parameters for h$^{-}$h$^{-}$ and identified $\pi^{-}\pi^{-}$
pairs. To enable the particle identification by specific energy loss in the TPC the particle momenta in the laboratory
were required to fall into the interval from 5 to 50~GeV/c. This selection allows a comparison only in certain
kinematic regions. Where the comparison is possible good agreement was found. An example is given in 
Fig.~\ref{fig:radii_withandwithout_dedx}, where the $k_{t}$ dependence of fit parameters at midrapidity 
is shown at 158{\it A} GeV beam energy from analyses with and without particle identification (PID),
in the latter case with corrections according to Eq.~\ref{equation:finalfit}.
\begin{figure}[h]
\centering
\resizebox{0.45\textwidth}{!}{
\includegraphics{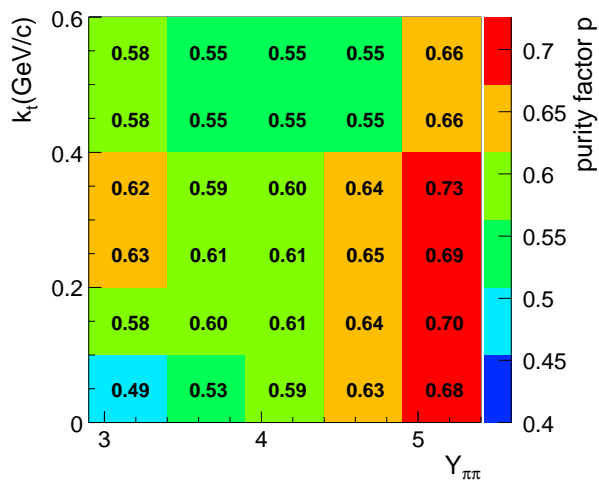}}
\caption{(Color online)~The purity factor $p$ as function of  $Y_{\pi\pi}$ and $k_{t}$ derived from simulations at 158{\it A} GeV beam energy.}
\label{fig:purity}
\end{figure} 

\begin{figure}[h]
\centering
\resizebox{0.45\textwidth}{!}{
\includegraphics{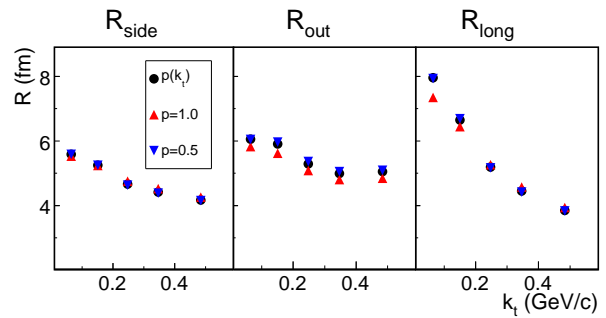}}
\caption{\label{fig:purity_radii} (Color online)~$k_{t}$ dependence of fit parameters at midrapidity at 158{\it A} GeV assuming different 
        values for the purity factor.}
\end{figure} 

The measured track sample is not only contaminated with non-pions. Also weak decays of heavier particles away 
from the vertex can generate pions which cannot be distinguished experimentally from primary particles. 
These decay products show no
correlations with pions generated in the interaction volume. Their contribution to the total number of 
pion pairs is estimated from simulations and also included in the purity factor $p$. Fig.~\ref{fig:purity} shows 
the purity $p$ in the analyzed $Y_{\pi\pi}$~,~$k_{t}$~ bins at 158{\it A} GeV beam energy.

Fig.~\ref{fig:purity_radii} demonstrates that the fitted radii change only little when 
instead of the purity factors shown in Fig.~\ref{fig:purity} fixed values of 0.5 and 1.0 are used.
In the following analyses the $Y_{\pi\pi}$~ and $k_{t}$ dependent purity estimates are used.

\subsection{Parametrization of the Coulomb interaction}
\label{sec:coulomb}

Identical particle correlations occur not only due to the quantum character of the particles. In addition,
electromagnetic repulsion between particles of like sign charge modifies the correlation function significantly at 
small momentum differences.
In order to extract the correlations due to Bose-Einstein statistics, the Coulomb interaction is taken 
into account during the fitting procedure.  According to \cite{Sinyukov:1998fc} the electromagnetic 
repulsion effect can be parametrized as a function of the modulus of the  
momentum difference $k^{\ast} = \left | \bold{k^{\ast}} \right |$ with
$\bold{k^{\ast}} = \bold{p_{1}^{\ast}} = - \bold{p_{2}^{\ast}}$ and the mean pair separation $\langle r^{\ast} \rangle$, where
the asterisk indicates values calculated in the c.m.s. of the particle pair.
If the momentum difference is smaller than $\tilde{k}$, defined as 

\begin{equation}
\label{eq:coulomb_low}
\tilde{k} = \frac{\pi}{4 \langle r^{\ast} \rangle } \left[ 1 + 2 \frac{\langle r^{\ast} \rangle}{a} \left( 1 +d_{2}  \frac{\langle r^{\ast} \rangle}{a} \right) \right],
\end{equation} 

\noindent with the $\pi^{-}\pi^{-}$ Bohr radius $a = 388$~fm and  $d_{2} = 3\pi / 8$, the Coulomb weight is given by ($k^{\ast}<\tilde{k}$):

\begin{equation}
C_{Coulomb}(\eta) = A_{C}(\eta) \left[ 1 + 2 \frac{\langle r^{\ast} \rangle}{a} \left( 1 +d_{2}  \frac{\langle r^{\ast} \rangle}{a}
\right) \right],
\end{equation}

\noindent with the Gamov term $A_{C}(\eta)=2\pi\eta(\exp(2\pi\eta)-1)^{-1}$ and  $\eta = 1/(k^{\ast}a)$.
In case $k^{\ast}$ is larger than $\tilde{k}$, the Coulomb weight is given by ($k^{\ast}>\tilde{k}$):

\begin{equation}
C_{Coulomb}(\eta) = \left( 1 - \frac{d}{a \langle r^{\ast} \rangle {k^{\ast}}^{2}} \right)
\end{equation}

\noindent with

\begin{equation}
\label{eq:coulomb_d}
d = a \langle r^{\ast} \rangle \tilde{k} 
                \left[ 1 - A_{C}(\eta)
\left[ 1 + 2 \frac{\langle r^{\ast} \rangle}{a} \left( 1 +d_{2}  \frac{\langle r^{\ast} \rangle}{a} \right) \right]
                \right].
\end{equation}

\noindent Under the assumption that both $\langle r^{\ast} \rangle$ and the BE parameters are fixed 
by the same freeze-out conditions and that
$R_{long} \simeq R_{side} \simeq R$ and $R_{out}^{\ast} = \frac{m_{t}}{m} R_{out}$ the mean 
pair separation $\langle r^{\ast} \rangle$ is approximated in \cite{Sinyukov:1998fc} by

\begin{equation}
\langle r^{\ast} \rangle \sim \frac{2}{\sqrt{\pi}} R_{out}^{\ast} 
\left[ 1+ (1-\epsilon^{2}) 
        \frac{1}{2\epsilon} ln{\frac{1+\epsilon}{1-\epsilon}} \right],
\label{eq:rstar}
\end{equation}

\noindent where $\epsilon = (1-(R/R_{out}^{\ast})^{2})^{1/2}$

During the fitting procedure $R$ is approximated by $R \sim (R_{long} + R_{side})/2$ and 
$\langle r^{\ast} \rangle$ is obtained iteratively from Eq.~\ref{eq:rstar}.

\begin{figure}[h]
\centering
\resizebox{0.45\textwidth}{!}{
\includegraphics{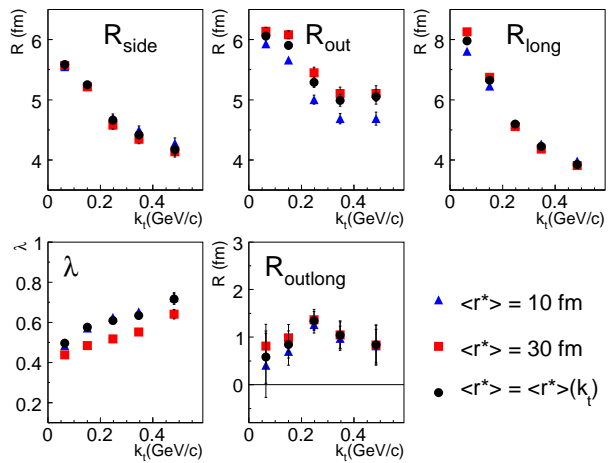}}
\caption{\label{fig:radii_at_diffcoulomb} (Color online)~Fit parameters (158$A$ GeV, midrapidity) using various 
values for the mean pair separation $\langle r^{\ast} \rangle$ in the Coulomb
correction procedure. Note the different scales on the the vertical axes.}
\end{figure}

Fig.~\ref{fig:radii_at_diffcoulomb} shows fitted BE parameters using Coulomb
weights based on $\langle r^{\ast} \rangle$ derived from the fitted  radii as well as results
obtained with fixed values of $\langle r^{\ast} \rangle$. Systematic variations of less than
5 \% are found in the radii for the different assumptions.

With the NA49 setup it is possible to measure the correlation function of oppositely charged particles. 
Based on this measurement a Coulomb correction procedure can be derived, not relying on any assumption about 
the mean pair separation. In this way the Coulomb interaction was taken into account in a previous NA49
publication \cite{Appelshauser:1997rr}.  Since in certain kinematic regions
the fraction of positively charged particles which do not contribute to the correlation function is only poorly known 
but certainly relatively large, the analytical procedure is preferred in the present analysis.

\subsection{Fitting procedure}
\label{sec:fitting}

The measured correlation function $C_{meas}(q)= S(q)/B(q)$ is fitted by the parametrization:

\begin{equation}
C_{meas}(q) = n \cdot ( p \cdot C_{BE} \cdot C_{Coulomb} + ( 1 - p ) ),
\label{eq:finalfitting}
\end{equation}

\noindent where $n$ represents a normalization factor introduced to account for the different number of entries
in the signal and reference distributions. The normalization can be derived in different ways. In this
analysis it is determined in the fitting procedure where it is treated as a free parameter.
The remaining terms of Eq.~\ref{eq:finalfitting}, the purity $p$, the Bose-Einstein correlation $C_{BE}$,
and the Coulomb correlation $C_{Coulomb}$ have been introduced in the preceding paragraphs. 

The free parameters, i.e. the radii $R_{side}$, $R_{out}$, $R_{long}$, $R_{outlong}$, the
incoherence parameter $\lambda$ and the normalisation constant $n$ are
obtained from a minimization of

\begin{eqnarray}
\chi^2 &=& \sum_{i} ( S(q_i)/B(q_i) - C_{meas}(q_i) )^2 / e(q_i)^2 \\
 e(q_i)   &=&   S(q_i)/B(q_i) \cdot  \sqrt{1/S(q_i) + 1/B(q_i)} \nonumber 
\end{eqnarray}

\noindent using the MINUIT package \cite{James:1994}.
The fitting routine includes all bins $i$ with $|\bold{q}|<0.2$~GeV/c. Varying the fitted range
changed the resulting fit parameters by less than 5\%.
The MINUIT package attributes an error to each parameter, which is equal to the inverse of the second
derivative of the $\chi^2$ function at its minimum with respect to this parameter. 
Along with the numerical value of each parameter at the minimum of the $\chi^2$ function
its uncertainty is always given.  
A quantitative estimate of the fit quality is not given, since it depends strongly on the fitted range
and details of the fitting procedure. However,
comparing the projections of the fitted and measured correlation functions
in Fig.~\ref{fig:Projections}, it can be concluded that Eq.~\ref{eq:finalfitting}
describes the data very well.

\subsection{Systematic uncertainties}
The uncertainty of the extracted BE parameters is characterized not only by the error estimated 
during the fitting procedure, but also by the systematic uncertainties in the whole analysis procedure.
Indeed, due to the large statistics of the pair samples,
the systematic errors are expected to be larger than the statistical ones.

In order to estimate the systematic errors, the crucial analysis cuts are varied over a reasonable range and the impact of
different sets of cuts on the fit parameters is calculated. The following paragraphs discuss the individual cuts which were
varied and the impact on the values of the radii.  

A detailed comparison to the analysis presented in \cite{Adamova:2002wi} revealed that the way of normalizing the correlation function
may be relevant. In this analysis the normalization is treated as a free parameter in the fit procedure, but it 
can also be derived from the ratio of the number of entries in the signal distribution to the number of entries in the background distribution.
The difference between the fit parameters determined using the two different approaches is taken as an estimate for the systematic
uncertainty due to the normalization method. On average, the systematic uncertainty due to the unknown normalization amounts to 
about 0.1 fm in the radii.

The way the Coulomb interaction is taken into account influences the final results. The size of this effect is estimated by
assuming different values for the mean pair separation in Eqs.~\ref{eq:coulomb_low}~-~\ref{eq:coulomb_d}. The fit is done assuming
$\langle r^{\ast} \rangle = 10$~fm and $\langle r^{\ast} \rangle = 30$~fm and the resulting parameters are compared to the standard 
procedure where $\langle r^{\ast} \rangle$ is determined iteratively from the radius parameters. The maximum difference between the 
standard procedure and the two extreme cases is taken as an estimate of the systematic uncertainty due to the treatment of the Coulomb interaction.
Also this uncertainty may change the final values of the radii by the order of 0.1 fm.  

The determination of the purity factor may also influence the final results in a systematic way. Hence the fit was done assuming a purity
20\% larger and 20\% smaller than the purity derived from the simulations. Again, the maximum difference between the standard approach and the
extreme values is taken as an estimate or the uncertainty introduced by the purity determination and amounts in most cases to less than
0.1 fm.

One of the most sensitive pair cuts is the anti-merging cut described in paragraph 
\ref{chap:Two track resolution cut}. As can be seen from Fig.~\ref{fig:routvsdcacut} 
the fit results change considerably with the number of required rows and the minimum separation. Varying the cut values in 
a reasonable range (45 $<$ number of rows $<$ 72 and 1.8 cm $<$ minimum separation $<$ 3.0 cm) and comparing the fit results to the standard cut
(number of rows = 50 and  minimum separation = 2.2 cm) the size of this systematic uncertainty is estimated.  Since it is very time-consuming
this procedure is only performed for the bin with the largest expected uncertainty, i.e. at high transverse momentum, at midrapidity 
and at 158{\it A} GeV beam energy. The systematic error due to this uncertainty in the reconstruction process 
reaches values up to 0.5 fm.

The overall systematic error assigned to the radius parameters is defined as the maximum value of the four errors described above. 
For the $\lambda$ parameter the variation of the purity value was not considered, since these two parameters are directly correlated.
In Table~\ref{tab:results} both the error from statistics and the systematic error are given. The 
figures in section~\ref{chap:res} show statistical errors only.

\begin{figure}[htb!]
\centering
\resizebox{0.45\textwidth}{!}{
\includegraphics{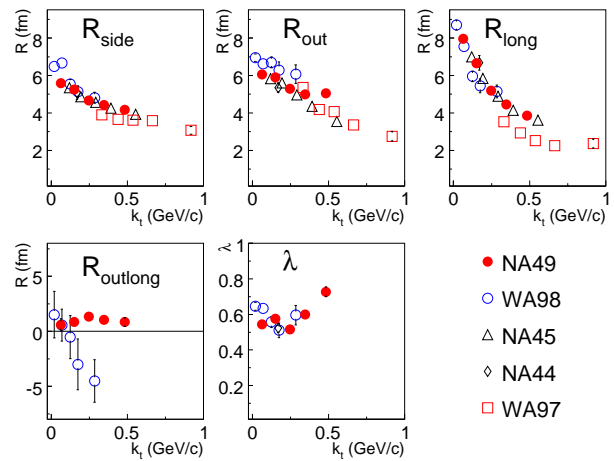}}
\caption{Dependence of fit parameters on $k_{t}$ at midrapidity at 158{\it A} GeV beam energy. 
Open symbols indicate measurements by other experiments 
\cite{Bearden:1998aq,Antinori:2001yi,Rosselet:2002ru,Adamova:2002wi}.}
\label{fig:ktdep_158_midrap}
\end{figure} 

\section{Results}
\label{chap:res}
\subsection{Energy and $\boldsymbol{k_{t}}$ dependence of fit parameters at midrapidity}

Fig.~\ref{fig:ktdep_158_midrap} shows the $k_{t}$ dependence of the extracted BE parameters 
at midrapidity ($2.9 < Y_{\pi\pi} < 3.4$) at the top SPS beam energy of 158$A$ GeV from
this analysis (filled dots) along with previously published results from 
experiments NA44 \cite{Bearden:1998aq}, WA97 \cite{Antinori:2001yi}, 
WA98 \cite{Rosselet:2002ru}, and NA45 \cite{Adamova:2002wi}.
  The parameters $R_{side}$, $R_{out}$, and $R_{long}$ decrease with increasing transverse momentum. 
In heavy-ion collisions, this behaviour is commonly attributed to collective expansion, with the rate of decrease 
reflecting the strength of the expansion. In section~\ref{chap:discussion}, the $k_{t}$ dependence will be compared to model calculations.
At vanishing transverse momentum $R_{out}$ and $R_{side}$ are expected to converge. Indeed the 
increase of $R_{out}$ flattens at low $k_{t}$ and the values of $R_{out}$ approach those of $R_{side}$.
The parameter $R_{outlong}$ remains small, independent of transverse momentum, as expected at midrapidity. 

The $\lambda$ parameter deviates from unity. Since the dilution of the primary pion pairs by secondary pions
and non-pions is already included via the purity factor,
the departure of the $\lambda$ parameter from unity has to be attributed to physics effects.
One often proposed explanation is the incomplete chaoticity of the source.
The value of the $\lambda$ parameter could also be related to features of the pion emission,
e.g. \cite{Gavrilik:2005zi} discusses the impact of the non-zero proper volume of the pions 
on the correlation function.
It could also be partially caused by a non-Gaussian shape of the
correlation function due to e.g. long-lived resonance states, or the approximation used
to parametrize the Coulomb interaction \cite{Kisiel:2006is}.
Details of the functional form of the correlation function are
currently studied using source imaging methods \cite{Chung:2007si}
introduced in \cite{Brown:1997ku,Brown:1997sn}.

Fig.~\ref{fig:ktdep_158_midrap} also displays published results from other experiments 
\cite{Bearden:1998aq,Antinori:2001yi,Rosselet:2002ru,Adamova:2002wi} for comparison.
One should note that the analysis procedures applied by the collaborations differ in details, 
e.g. in the treatment of the Coulomb interaction. Moreover,
some experiments did not use the mean transverse momentum variable $k_{t}$, but some similar measure. 
In Fig.~\ref{fig:ktdep_158_midrap} differences between those kinematic variables are neglected. 
Absolute values as well as transverse momentum dependence of radii extracted by the various
experiments are similar.

\begin{figure}[htb]
\centering
\resizebox{0.45\textwidth}{!}{
\includegraphics{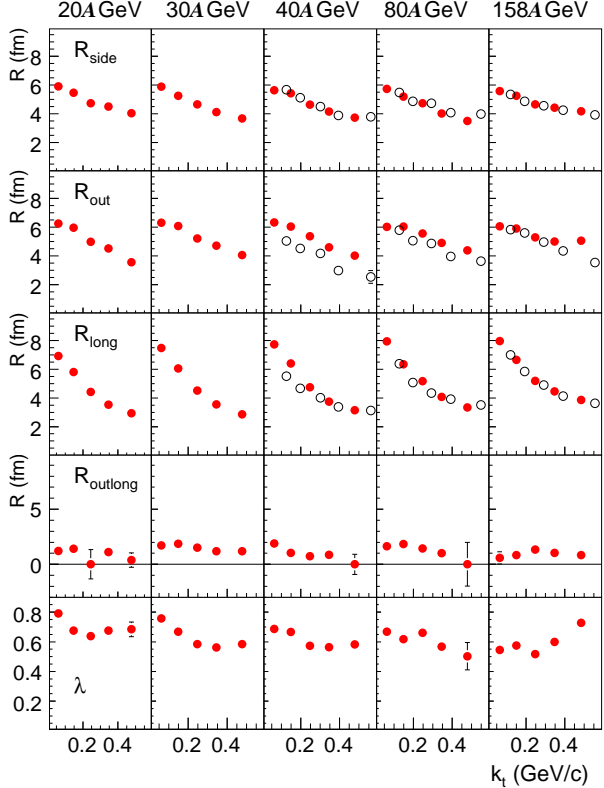}}
\caption{Dependence of fit parameters on $k_{t}$ at midrapidity for different beam energies. Full dots 
show results from this analysis, open circles indicate measurements of NA45 \cite{Adamova:2002wi}.}
\label{fig:ktdep_5energies_midrap}
\end{figure}

In Fig.~\ref{fig:ktdep_5energies_midrap} the $k_{t}$ dependence of the fit parameters is plotted
at five different beam energies in the central rapidity bin. At all energies, the transverse momentum
dependence of the fit parameters is similar. The radii decrease with increasing transverse 
momentum, neither absolute values nor gradients change significantly with increasing beam energy. 
The cross term $R_{outlong}$ shows small values independent of energy and transverse momentum.
The $\lambda$ parameter decreases weakly with increasing beam energy. 

At 40$A$ and 80$A$ and 158$A$~GeV beam energy, measurements of NA45 \cite{Adamova:2002wi} are also shown. 
The radii agree in the side and long component, but NA45 reports systematically smaller values of $R_{out}$. 
The largest discrepancy is observed at 40$A$~GeV beam energy. A detailed comparison study excluded that the origin of the
difference lies in the treatment of the Coulomb interaction or in the fitting procedure.
Rather it is due to differences in the measured correlation functions themselves.

\begin{figure}[h!]
\centering
\resizebox{0.45\textwidth}{!}{
\includegraphics{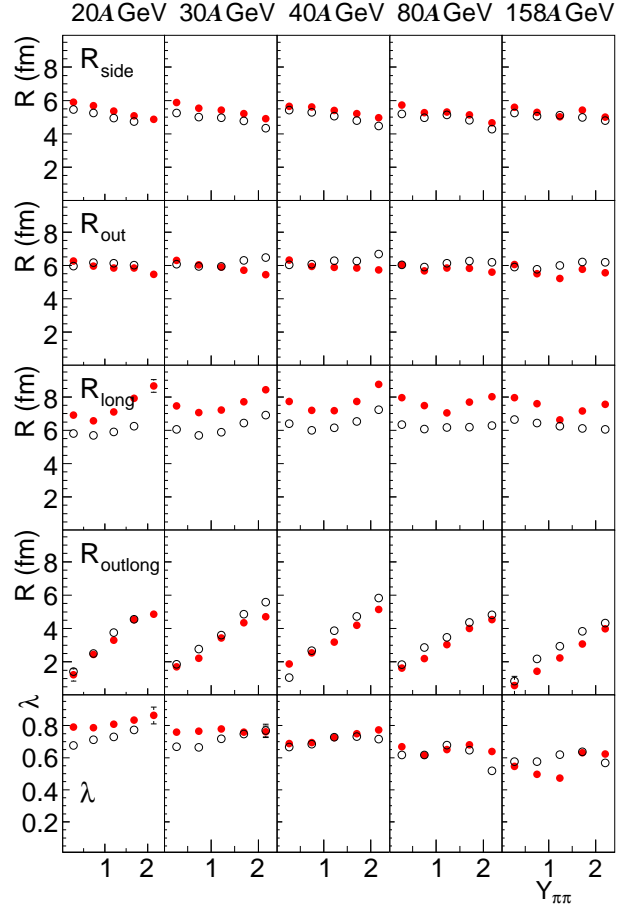}}
\caption{Dependence of the fit parameters on rapidity at low $k_{t}$ at different beam energies. Filled symbols 
correspond to $0.0\!<\!k_{t}\!<0.1\!$~GeV/c,
open symbols to ${0.1<k_{t}<0.2}$~GeV/c. 
\label{fig:rapdep_5energies1}}
\end{figure} 

\begin{figure}[h!]
\centering
\resizebox{0.45\textwidth}{!}{
\includegraphics{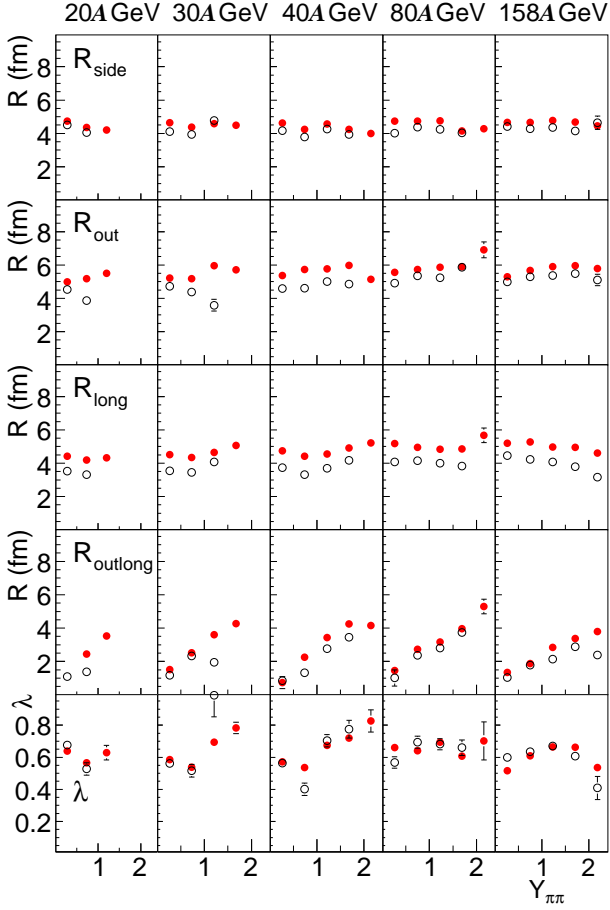}}
\caption{Dependence of the fit parameters on rapidity at high $k_{t}$ at different beam energies. Filled symbols 
correspond to ${0.2<k_{t}<0.3}$~GeV/c, open symbols to $0.3<k_{t}<0.4$~GeV/c.}
\label{fig:rapdep_5energies2}
\end{figure} 

\subsection{Dependence of fit parameters on rapidity}
The dependence of the fit parameters on rapidity is shown in Figs.~\ref{fig:rapdep_5energies1} and~\ref{fig:rapdep_5energies2} 
for the 5 beam energies of this analysis. 
At lower transverse momenta (see Fig.~\ref{fig:rapdep_5energies1}), the parameter $R_{side}$ decreases weakly with 
increasing rapidity, with a larger gradient at smaller beam energies.
The parameter $R_{out}$ remains almost constant at all rapidities and at all energies.
The rapidity dependence of the parameter $R_{long}$ shows no clear systematic trend. At small transverse momenta ($0.0<k_{t}<0.1$~GeV/c) 
variations of the order of one fermi are observed.
The parameter $R_{outlong}$ increases clearly with increasing rapidity, as expected for particle sources with
non-vanishing longitudinal expansion \cite{Chapman:1995vq}.
The $\lambda$ parameter shows no distinct rapidity dependence, but it decreases with increasing energy from
values of about 0.8 at 20$A$ GeV beam energy to about 0.6 at 158$A$ GeV beam energy.

The rapidity dependence of the fit parameters at larger transverse momenta is displayed in Fig.~\ref{fig:rapdep_5energies2}.
Since the pion yield decreases with increasing transverse momentum and with increasing rapidity, the statistics in several
kinematic bins, especially at low beam energies at larger rapidities, is too small to obtain a reasonable fit result. 
The parameters $R_{side}$, $R_{out}$, and at larger transverse momenta also $R_{long}$, depend only weakly on rapidity and 
on beam energy. The parameter $R_{outlong}$ increases with increasing rapidity also at larger transverse momenta. 
The $\lambda$ parameter shows larger fluctuations in the case of  higher transverse momenta, but there is no obvious pattern 
visible.

\subsection{Numerical values}
Tables~\ref{tab:results} and~\ref{tab:results1} list numerical values of all fit parameters. The first three columns indicate beam energy, rapidity bin
and mean transverse momentum respectively. The next columns give the numerical values of the fitted parameters
with their statistical and systematic error 
in parentheses. For completeness the last two columns give the estimated purity factor and the 
mean pair separation used for the Coulomb correction and calculated iteratively from the fitted 
radius parameters. 

\begin{table*}[!htb]
\caption{\label{tab:results} Numerical results of the fitting procedure at 20${\it A}$, 30${\it A}$, and 40${\it A}$ GeV beam energy. 
        The first three columns give the beam energy, the rapidity interval,
        and the mean transverse momentum, respectively. The next five columns show the BE parameters, in the last two columns the purity
        factor and the assumed mean pair separation applied in the Coulomb correction procedure are listed.}
\centering
\resizebox{16.5cm}{!}
{
\begin{tabular}{cccccccccc}
\hline\noalign{\smallskip}
$E_{Beam}$  &  $Y_{\pi\pi}$  &  $\langle k_{t}\rangle$  &  $R_{side}$
&  $R_{out}$  &  $R_{long}$  &  $R_{outlong}$  &  $\lambda$ & $p$ & $ \langle r^{\ast} \rangle $\\
$({\it A}GeV)$  &    &  (GeV/c)  &  (fm)   &  (fm)  &  (fm)  &  (fm)   &   &   & (fm) \\

\noalign{\smallskip}\hline\noalign{\smallskip}

20 & 0.0-0.5 & 0.06 & 5.90(0.09)(0.16) & 6.25(0.09)(0.20) & 6.91(0.11)(0.09) & 1.22(0.38)(0.08) & 0.79(0.02)(0.15) & 0.39 &  14.76    \\
20 & 0.0-0.5 & 0.15 & 5.46(0.08)(0.13) & 5.95(0.08)(0.22) & 5.80(0.08)(0.23) & 1.41(0.25)(0.08) & 0.68(0.01)(0.13) & 0.51 &  15.17    \\
20 & 0.0-0.5 & 0.25 & 4.73(0.08)(0.14) & 4.98(0.10)(0.35) & 4.42(0.08)(0.24) & 0.00(1.34)(0.00) & 0.64(0.02)(0.09) & 0.63 &  15.12    \\
20 & 0.0-0.5 & 0.34 & 4.51(0.12)(0.35) & 4.52(0.14)(0.49) & 3.53(0.10)(0.32) & 1.10(0.28)(0.11) & 0.68(0.03)(0.07) & 0.67 &  16.43    \\
20 & 0.0-0.5 & 0.48 & 4.04(0.14)(0.79) & 3.56(0.16)(0.89) & 2.94(0.14)(0.83) & 0.37(0.65)(0.74) & 0.68(0.05)(0.18) & 0.65 &  16.50    \\
\cline{2-10}
20 & 0.5-1.0 & 0.06 & 5.70(0.08)(0.13) & 5.95(0.08)(0.20) & 6.56(0.10)(0.09) & 2.45(0.16)(0.16) & 0.79(0.01)(0.14) & 0.45 &  14.12   \\
20 & 0.5-1.0 & 0.15 & 5.24(0.09)(0.13) & 6.17(0.11)(0.25) & 5.69(0.10)(0.23) & 2.51(0.18)(0.18) & 0.71(0.02)(0.11) & 0.53 &  15.23    \\
20 & 0.5-1.0 & 0.25 & 4.36(0.10)(0.20) & 5.18(0.14)(0.35) & 4.19(0.11)(0.25) & 2.44(0.17)(0.17) & 0.56(0.02)(0.07) & 0.62 &  15.14    \\
20 & 0.5-1.0 & 0.34 & 4.05(0.16)(0.59) & 3.87(0.18)(0.75) & 3.31(0.14)(0.50) & 1.38(0.24)(0.19) & 0.53(0.04)(0.09) & 0.66 &  14.32    \\
\cline{2-10}
20 & 1.0-1.5 & 0.06 & 5.37(0.08)(0.13) & 5.85(0.09)(0.20) & 7.10(0.12)(0.13) & 3.30(0.14)(0.22) & 0.81(0.02)(0.14) & 0.51 &  14.20    \\
20 & 1.0-1.5 & 0.15 & 4.95(0.10)(0.15) & 6.14(0.13)(0.22) & 5.90(0.13)(0.23) & 3.76(0.16)(0.27) & 0.73(0.02)(0.12) & 0.57 &  15.11    \\
20 & 1.0-1.5 & 0.24 & 4.20(0.15)(0.35) & 5.50(0.25)(0.75) & 4.32(0.17)(0.46) & 3.54(0.23)(0.52) & 0.63(0.04)(0.05) & 0.65 &  15.73    \\
\cline{2-10}
20 & 1.5-2.0 & 0.06 & 5.08(0.11)(0.15) & 5.85(0.13)(0.20) & 7.93(0.18)(0.28) & 4.54(0.17)(0.32) & 0.83(0.02)(0.13) & 0.58 &  14.41    \\
20 & 1.5-2.0 & 0.14 & 4.73(0.11)(0.25) & 6.02(0.15)(0.37) & 6.24(0.16)(0.37) & 4.54(0.17)(0.31) & 0.77(0.03)(0.12) & 0.64 &  15.00    \\
\cline{2-10}
20 & 2.0-2.5 & 0.06 & 4.87(0.19)(0.52) & 5.46(0.23)(0.66) & 8.65(0.38)(1.21) & 4.86(0.32)(0.73) & 0.86(0.05)(0.11) & 0.67 &  13.40    \\
\cline{2-10}
\hline
30 & 0.0-0.5 & 0.06 & 5.88(0.08)(0.16) & 6.30(0.08)(0.20) & 7.47(0.10)(0.13) & 1.71(0.24)(0.15) & 0.76(0.01)(0.14) & 0.41 &  15.22   \\
30 & 0.0-0.5 & 0.15 & 5.25(0.07)(0.13) & 6.07(0.08)(0.25) & 6.05(0.08)(0.23) & 1.86(0.18)(0.13) & 0.67(0.01)(0.12) & 0.51 &  15.37   \\
30 & 0.0-0.5 & 0.25 & 4.65(0.08)(0.14) & 5.21(0.10)(0.35) & 4.51(0.08)(0.24) & 1.51(0.18)(0.08) & 0.58(0.02)(0.08) & 0.59 &  15.60    \\
30 & 0.0-0.5 & 0.35 & 4.11(0.10)(0.27) & 4.72(0.12)(0.36) & 3.55(0.10)(0.30) & 1.18(0.23)(0.08) & 0.56(0.02)(0.06) & 0.66 &  16.77    \\
30 & 0.0-0.5 & 0.48 & 3.66(0.11)(0.48) & 4.05(0.15)(0.75) & 2.85(0.09)(0.42) & 1.18(0.20)(2.35) & 0.58(0.03)(0.09) & 0.64 &  18.15    \\
\cline{2-10}
30 & 0.5-1.0 & 0.06 & 5.54(0.06)(0.13) & 6.03(0.07)(0.20) & 7.06(0.09)(0.10) & 2.21(0.16)(0.18) & 0.77(0.01)(0.14) & 0.45 &  14.45    \\
30 & 0.5-1.0 & 0.15 & 5.00(0.07)(0.13) & 5.95(0.09)(0.24) & 5.69(0.08)(0.23) & 2.77(0.13)(0.23) & 0.66(0.01)(0.11) & 0.53 &  14.79    \\
30 & 0.5-1.0 & 0.25 & 4.37(0.10)(0.14) & 5.17(0.13)(0.35) & 4.34(0.11)(0.24) & 2.52(0.16)(0.17) & 0.54(0.02)(0.06) & 0.60 &  15.23   \\
30 & 0.5-1.0 & 0.35 & 3.93(0.13)(0.36) & 4.38(0.21)(0.78) & 3.45(0.13)(0.47) & 2.32(0.19)(0.47) & 0.52(0.04)(0.08) & 0.66 &  15.69    \\
30 & 0.5-1.0 & 0.48 & 3.36(0.19)(1.00) & 3.68(0.26)(1.40) & 2.42(0.15)(1.05) & 1.36(0.24)(0.76) & 0.38(0.05)(0.52) & 0.66 &  16.41    \\
\cline{2-10}
30 & 1.0-1.5 & 0.06 & 5.43(0.06)(0.13) & 5.90(0.07)(0.20) & 7.21(0.09)(0.13) & 3.43(0.11)(0.26) & 0.78(0.01)(0.13) & 0.51 &  14.37    \\
30 & 1.0-1.5 & 0.15 & 4.97(0.06)(0.13) & 5.94(0.08)(0.22) & 5.88(0.09)(0.23) & 3.61(0.10)(0.29) & 0.72(0.01)(0.12) & 0.55 &  14.86   \\
30 & 1.0-1.5 & 0.24 & 4.59(0.12)(0.16) & 5.96(0.18)(0.35) & 4.66(0.14)(0.24) & 3.60(0.18)(0.24) & 0.69(0.03)(0.11) & 0.62 &  17.06    \\
30 & 1.0-1.5 & 0.34 & 4.77(0.23)(0.61) & 3.59(0.34)(1.72) & 4.07(0.20)(0.45) & 1.95(0.31)(0.94) & 0.99(0.13)(0.05) & 0.66 &  14.43    \\
\cline{2-10}
30 & 1.5-2.0 & 0.06 & 5.21(0.08)(0.13) & 5.71(0.09)(0.20) & 7.71(0.13)(0.26) & 4.35(0.12)(0.32) & 0.76(0.02)(0.12) & 0.59 &  14.12    \\
30 & 1.5-2.0 & 0.14 & 4.77(0.07)(0.13) & 6.30(0.10)(0.27) & 6.43(0.11)(0.23) & 4.86(0.11)(0.38) & 0.75(0.02)(0.11) & 0.63 &  15.51    \\
30 & 1.5-2.0 & 0.24 & 4.48(0.13)(0.32) & 5.71(0.19)(0.58) & 5.07(0.16)(0.46) & 4.26(0.18)(0.44) & 0.78(0.04)(0.09) & 0.68 &  16.71    \\
\cline{2-10}
30 & 2.0-2.5 & 0.06 & 4.91(0.14)(0.24) & 5.45(0.17)(0.32) & 8.43(0.28)(0.59) & 4.72(0.23)(0.40) & 0.76(0.04)(0.09) & 0.69 &  13.42    \\
30 & 2.0-2.5 & 0.14 & 4.34(0.13)(0.34) & 6.46(0.21)(0.58) & 6.91(0.25)(0.71) & 5.58(0.24)(0.57) & 0.77(0.04)(0.10) & 0.70 &  15.67    \\
\cline{2-10}
\hline
40 & 0.0-0.5 & 0.06 & 5.64(0.06)(0.15) & 6.32(0.07)(0.20) & 7.72(0.09)(0.25) & 1.88(0.20)(0.55) & 0.69(0.01)(0.13) & 0.43 &  15.26    \\
40 & 0.0-0.5 & 0.15 & 5.43(0.06)(0.13) & 6.04(0.07)(0.25) & 6.41(0.07)(0.23) & 1.05(0.28)(0.07) & 0.67(0.01)(0.12) & 0.53 &  15.70    \\
40 & 0.0-0.5 & 0.25 & 4.63(0.07)(0.14) & 5.37(0.09)(0.35) & 4.74(0.08)(0.24) & 0.73(0.37)(0.06) & 0.57(0.01)(0.08) & 0.58 &  16.03    \\
40 & 0.0-0.5 & 0.35 & 4.16(0.09)(0.20) & 4.59(0.12)(0.30) & 3.73(0.09)(0.22) & 0.85(0.30)(0.13) & 0.56(0.02)(0.06) & 0.62 &  16.54    \\
40 & 0.0-0.5 & 0.48 & 3.73(0.10)(0.36) & 4.02(0.14)(0.54) & 3.15(0.11)(0.36) & 0.00(0.92)(0.50) & 0.58(0.03)(0.08) & 0.62 &  18.20    \\
\cline{2-10}
40 & 0.5-1.0 & 0.06 & 5.60(0.06)(0.13) & 5.95(0.06)(0.20) & 7.19(0.08)(0.16) & 2.54(0.12)(0.22) & 0.69(0.01)(0.13) & 0.47 &  14.53    \\
40 & 0.5-1.0 & 0.15 & 5.30(0.05)(0.13) & 6.08(0.07)(0.23) & 6.00(0.07)(0.23) & 2.67(0.11)(0.25) & 0.68(0.01)(0.12) & 0.54 &  15.37    \\
40 & 0.5-1.0 & 0.25 & 4.25(0.09)(0.14) & 5.72(0.14)(0.40) & 4.42(0.09)(0.24) & 2.24(0.18)(0.22) & 0.54(0.02)(0.08) & 0.57 &  16.30    \\
40 & 0.5-1.0 & 0.35 & 3.79(0.14)(0.27) & 4.60(0.27)(0.73) & 3.31(0.15)(0.30) & 1.32(0.32)(0.28) & 0.40(0.04)(0.05) & 0.61 &  16.16    \\
40 & 0.5-1.0 & 0.48 & 3.60(0.20)(0.56) & 5.01(0.43)(1.68) & 2.73(0.17)(0.40) & 1.62(0.30)(0.45) & 0.48(0.07)(0.11) & 0.63 &  21.67    \\
\cline{2-10}
40 & 1.0-1.5 & 0.06 & 5.41(0.05)(0.13) & 5.88(0.06)(0.20) & 7.17(0.07)(0.14) & 3.18(0.09)(0.26) & 0.73(0.01)(0.13) & 0.52 &  14.32    \\
40 & 1.0-1.5 & 0.15 & 5.06(0.05)(0.13) & 6.29(0.07)(0.26) & 6.15(0.07)(0.23) & 3.87(0.08)(0.34) & 0.73(0.01)(0.12) & 0.55 &  15.56    \\
40 & 1.0-1.5 & 0.24 & 4.56(0.08)(0.14) & 5.77(0.11)(0.35) & 4.56(0.09)(0.24) & 3.43(0.11)(0.25) & 0.67(0.02)(0.11) & 0.57 &  16.61    \\
40 & 1.0-1.5 & 0.34 & 4.26(0.16)(0.38) & 5.00(0.20)(0.58) & 3.69(0.13)(0.34) & 2.77(0.19)(0.37) & 0.70(0.04)(0.09) & 0.59 &  17.66    \\
\cline{2-10}
40 & 1.5-2.0 & 0.06 & 5.22(0.06)(0.13) & 5.83(0.07)(0.20) & 7.72(0.10)(0.27) & 4.19(0.10)(0.35) & 0.75(0.01)(0.12) & 0.59 &  14.44    \\
40 & 1.5-2.0 & 0.15 & 4.80(0.06)(0.13) & 6.25(0.08)(0.30) & 6.53(0.08)(0.23) & 4.73(0.09)(0.40) & 0.73(0.01)(0.11) & 0.62 &  15.56    \\
40 & 1.5-2.0 & 0.24 & 4.25(0.08)(0.16) & 5.97(0.12)(0.35) & 4.92(0.11)(0.24) & 4.24(0.12)(0.30) & 0.72(0.02)(0.09) & 0.65 &  16.98    \\
40 & 1.5-2.0 & 0.34 & 3.93(0.16)(0.69) & 4.86(0.24)(1.48) & 4.18(0.18)(0.91) & 3.45(0.21)(1.05) & 0.77(0.06)(0.15) & 0.66 &  17.32    \\
\cline{2-10}
40 & 2.0-2.5 & 0.06 & 4.97(0.10)(0.13) & 5.72(0.12)(0.20) & 8.75(0.21)(0.59) & 5.13(0.17)(0.50) & 0.77(0.03)(0.09) & 0.65 &  14.13    \\
40 & 2.0-2.5 & 0.14 & 4.48(0.10)(0.16) & 6.69(0.16)(0.39) & 7.23(0.17)(0.36) & 5.83(0.17)(0.50) & 0.72(0.03)(0.09) & 0.68 &  16.27    \\
40 & 2.0-2.5 & 0.24 & 3.99(0.16)(0.53) & 5.14(0.26)(1.30) & 5.21(0.24)(0.93) & 4.16(0.23)(1.04) & 0.83(0.07)(0.07) & 0.69 &  15.32    \\
\cline{2-10}
\hline
\noalign{\smallskip}
\end{tabular}
}

\end{table*}

\begin{table*} [!htb]
\caption{\label{tab:results1} Numerical results of the fitting procedure at 80${\it A}$ and 158${\it A}$ GeV beam energy. 
        The first three columns give the beam energy, the rapidity interval,
        and the mean transverse momentum, respectively. The next five columns show the BE parameters, in the last two columns the purity
        factor and the assumed mean pair separation applied in the Coulomb correction procedure are listed.}
\centering
\resizebox{16.5cm}{!}
{
\centering
\begin{tabular}{cccccccccc}
\hline\noalign{\smallskip}
$Beam $  &  $Y_{\pi\pi}$  &  $\langle k_{t}\rangle$  &  $R_{side}$  &  $R_{out}$  &  $R_{long}$  &  $R_{outlong}$  &  $\lambda$ & $p$ & $ \langle r^{\ast} \rangle $\\
$({\it A}GeV)$  &    &  (GeV/c)  &  (fm)   &  (fm)  &  (fm)  &  (fm)   &   &   & (fm) \\
\noalign{\smallskip}\hline\noalign{\smallskip}

80 & 0.0-0.5 & 0.06 & 5.73(0.08)(0.13) & 6.03(0.08)(0.20) & 7.95(0.12)(0.37) & 1.63(0.29)(0.22) & 0.67(0.01)(0.11) & 0.45 &  14.94    \\
80 & 0.0-0.5 & 0.15 & 5.20(0.07)(0.13) & 6.03(0.09)(0.33) & 6.33(0.09)(0.23) & 1.84(0.21)(0.27) & 0.62(0.01)(0.10) & 0.56 &  15.52    \\
80 & 0.0-0.5 & 0.25 & 4.74(0.08)(0.14) & 5.55(0.11)(0.35) & 5.17(0.10)(0.24) & 1.45(0.25)(0.12) & 0.66(0.02)(0.10) & 0.62 &  16.73    \\
80 & 0.0-0.5 & 0.35 & 4.01(0.13)(0.21) & 4.91(0.18)(0.45) & 4.07(0.14)(0.29) & 1.01(0.48)(0.22) & 0.57(0.04)(0.07) & 0.62 &  17.51    \\
80 & 0.0-0.5 & 0.48 & 3.50(0.23)(0.66) & 4.39(0.41)(1.49) & 3.34(0.24)(0.71) & 0.00(1.99)(0.51) & 0.50(0.09)(0.22) & 0.54 &  19.59    \\
\cline{2-10}
80 & 0.5-1.0 & 0.06 & 5.27(0.07)(0.13) & 5.67(0.08)(0.20) & 7.49(0.11)(0.34) & 2.20(0.19)(0.28) & 0.61(0.01)(0.10) & 0.51 &  14.06    \\
80 & 0.5-1.0 & 0.15 & 4.97(0.06)(0.15) & 5.90(0.08)(0.28) & 6.07(0.08)(0.23) & 2.87(0.12)(0.32) & 0.62(0.01)(0.11) & 0.57 &  15.01    \\
80 & 0.5-1.0 & 0.25 & 4.75(0.09)(0.15) & 5.72(0.11)(0.41) & 4.95(0.10)(0.24) & 2.73(0.15)(0.32) & 0.64(0.02)(0.11) & 0.60 &  16.92    \\
80 & 0.5-1.0 & 0.35 & 4.38(0.13)(0.23) & 5.35(0.17)(0.37) & 4.14(0.14)(0.26) & 2.37(0.23)(0.25) & 0.69(0.04)(0.10) & 0.60 &  18.93    \\
80 & 0.5-1.0 & 0.48 & 3.87(0.18)(0.63) & 5.05(0.24)(0.84) & 2.94(0.14)(0.54) & 1.69(0.29)(0.33) & 0.63(0.05)(0.09) & 0.49 &  21.99    \\
\cline{2-10}
80 & 1.0-1.5 & 0.06 & 5.30(0.07)(0.13) & 5.84(0.08)(0.20) & 7.04(0.10)(0.21) & 3.03(0.13)(0.32) & 0.65(0.01)(0.11) & 0.54 &  14.11    \\
80 & 1.0-1.5 & 0.15 & 5.13(0.06)(0.13) & 6.13(0.08)(0.30) & 6.16(0.08)(0.23) & 3.46(0.10)(0.35) & 0.68(0.01)(0.12) & 0.59 &  15.47    \\
80 & 1.0-1.5 & 0.25 & 4.75(0.08)(0.14) & 5.87(0.10)(0.39) & 4.83(0.09)(0.24) & 3.17(0.12)(0.29) & 0.69(0.02)(0.10) & 0.62 &  17.11    \\
80 & 1.0-1.5 & 0.34 & 4.24(0.12)(0.25) & 5.23(0.15)(0.36) & 3.99(0.13)(0.28) & 2.81(0.16)(0.23) & 0.68(0.03)(0.09) & 0.60 &  18.42    \\
80 & 1.0-1.5 & 0.47 & 3.41(0.13)(0.56) & 4.27(0.17)(0.88) & 2.73(0.12)(0.55) & 2.15(0.17)(0.43) & 0.55(0.03)(0.06) & 0.52 &  18.69    \\
\cline{2-10}
80 & 1.5-2.0 & 0.06 & 5.14(0.08)(0.13) & 5.83(0.09)(0.20) & 7.69(0.12)(0.31) & 3.99(0.12)(0.37) & 0.68(0.02)(0.11) & 0.61 &  14.47    \\
80 & 1.5-2.0 & 0.15 & 4.81(0.07)(0.13) & 6.27(0.09)(0.37) & 6.19(0.10)(0.23) & 4.37(0.10)(0.45) & 0.65(0.02)(0.10) & 0.64 &  15.43    \\
80 & 1.5-2.0 & 0.24 & 4.14(0.10)(0.14) & 5.90(0.15)(0.45) & 4.86(0.14)(0.24) & 3.97(0.15)(0.39) & 0.61(0.03)(0.07) & 0.66 &  16.80    \\
80 & 1.5-2.0 & 0.34 & 4.03(0.15)(0.48) & 5.85(0.24)(0.78) & 3.83(0.15)(0.41) & 3.74(0.19)(0.40) & 0.66(0.05)(0.06) & 0.63 &  19.89    \\
\cline{2-10}
80 & 2.0-2.5 & 0.06 & 4.66(0.11)(0.13) & 5.60(0.14)(0.20) & 8.01(0.21)(0.47) & 4.53(0.18)(0.45) & 0.64(0.03)(0.09) & 0.69 &  13.86    \\
80 & 2.0-2.5 & 0.15 & 4.28(0.13)(0.15) & 6.19(0.20)(0.48) & 6.27(0.22)(0.37) & 4.83(0.21)(0.59) & 0.52(0.03)(0.06) & 0.70 &  14.97    \\
80 & 2.0-2.5 & 0.24 & 4.29(0.24)(0.42) & 6.91(0.48)(1.27) & 5.68(0.44)(0.72) & 5.30(0.44)(0.86) & 0.70(0.12)(0.06) & 0.70 &  19.23    \\
\cline{2-10}
\hline
158 & 0.0-0.5 & 0.06 & 5.59(0.05)(0.13) & 6.06(0.06)(0.20) & 7.95(0.09)(0.63) & 0.58(0.55)(0.39) & 0.54(0.01)(0.08) & 0.49 &  15.06   \\
158 & 0.0-0.5 & 0.15 & 5.25(0.05)(0.13) & 5.90(0.06)(0.41) & 6.65(0.07)(0.28) & 0.85(0.29)(0.28) & 0.58(0.01)(0.07) & 0.58 &  15.64   \\
158 & 0.0-0.5 & 0.25 & 4.66(0.06)(0.14) & 5.29(0.08)(0.42) & 5.19(0.08)(0.24) & 1.34(0.20)(0.10) & 0.52(0.01)(0.09) & 0.63 &  16.21   \\
158 & 0.0-0.5 & 0.35 & 4.42(0.07)(0.15) & 4.99(0.10)(0.39) & 4.45(0.08)(0.16) & 1.04(0.28)(0.06) & 0.60(0.02)(0.10) & 0.62 &  18.20   \\
158 & 0.0-0.5 & 0.48 & 4.17(0.09)(0.15) & 5.05(0.12)(0.39) & 3.85(0.09)(0.15) & 0.84(0.41)(0.05) & 0.73(0.03)(0.10) & 0.58 &  22.80   \\
\cline{2-10}
158 & 0.5-1.0 & 0.06 & 5.30(0.08)(0.13) & 5.50(0.08)(0.22) & 7.59(0.12)(0.74) & 1.44(0.27)(0.29) & 0.50(0.02)(0.05) & 0.53 &  13.67   \\
158 & 0.5-1.0 & 0.15 & 5.06(0.05)(0.13) & 5.77(0.07)(0.37) & 6.44(0.07)(0.23) & 2.18(0.12)(0.26) & 0.58(0.01)(0.09) & 0.60 &  15.19   \\
158 & 0.5-1.0 & 0.25 & 4.67(0.05)(0.18) & 5.66(0.07)(0.44) & 5.27(0.06)(0.24) & 1.87(0.14)(0.20) & 0.61(0.01)(0.11) & 0.61 &  16.97   \\
158 & 0.5-1.0 & 0.35 & 4.29(0.07)(0.17) & 5.29(0.09)(0.41) & 4.22(0.07)(0.16) & 1.78(0.15)(0.10) & 0.63(0.02)(0.10) & 0.59 &  18.80   \\
158 & 0.5-1.0 & 0.48 & 4.08(0.08)(0.12) & 5.03(0.11)(0.41) & 3.61(0.08)(0.14) & 1.17(0.24)(0.03) & 0.72(0.03)(0.09) & 0.55 &  22.46   \\
\cline{2-10}
158 & 1.0-1.5 & 0.06 & 5.03(0.06)(0.13) & 5.21(0.06)(0.20) & 6.63(0.08)(0.36) & 2.24(0.11)(0.29) & 0.47(0.01)(0.06) & 0.59 &  12.95   \\
158 & 1.0-1.5 & 0.15 & 5.11(0.04)(0.13) & 6.00(0.05)(0.31) & 6.24(0.05)(0.23) & 2.94(0.08)(0.31) & 0.62(0.01)(0.11) & 0.61 &  15.37   \\
158 & 1.0-1.5 & 0.25 & 4.78(0.05)(0.15) & 5.89(0.06)(0.44) & 4.98(0.05)(0.24) & 2.84(0.08)(0.27) & 0.66(0.01)(0.11) & 0.61 &  17.31   \\
158 & 1.0-1.5 & 0.35 & 4.36(0.07)(0.14) & 5.37(0.08)(0.43) & 4.07(0.07)(0.13) & 2.14(0.12)(0.12) & 0.67(0.02)(0.09) & 0.60 &  18.96   \\
158 & 1.0-1.5 & 0.48 & 4.10(0.10)(0.21) & 4.98(0.12)(0.47) & 3.21(0.08)(0.16) & 1.68(0.16)(0.03) & 0.67(0.03)(0.06) & 0.55 &  21.99   \\
\cline{2-10}
158 & 1.5-2.0 & 0.06 & 5.42(0.05)(0.13) & 5.77(0.05)(0.20) & 7.16(0.07)(0.23) & 3.06(0.09)(0.33) & 0.63(0.01)(0.11) & 0.63 &  14.24   \\
158 & 1.5-2.0 & 0.15 & 4.98(0.04)(0.14) & 6.20(0.05)(0.33) & 6.12(0.06)(0.23) & 3.83(0.06)(0.39) & 0.64(0.01)(0.11) & 0.64 &  15.44   \\
158 & 1.5-2.0 & 0.25 & 4.68(0.06)(0.14) & 5.95(0.08)(0.47) & 4.95(0.07)(0.24) & 3.38(0.09)(0.29) & 0.66(0.02)(0.10) & 0.65 &  17.33   \\
158 & 1.5-2.0 & 0.34 & 4.15(0.11)(0.15) & 5.48(0.13)(0.54) & 3.80(0.12)(0.13) & 2.87(0.13)(0.18) & 0.61(0.03)(0.07) & 0.64 &  18.97   \\
158 & 1.5-2.0 & 0.47 & 3.86(0.17)(0.47) & 5.59(0.24)(0.83) & 3.13(0.15)(0.31) & 2.36(0.22)(0.08) & 0.69(0.07)(0.10) & 0.55 &  24.00   \\
\cline{2-10}
158 & 2.0-2.5 & 0.06 & 5.00(0.07)(0.13) & 5.55(0.07)(0.20) & 7.55(0.10)(0.34) & 3.98(0.09)(0.39) & 0.62(0.01)(0.10) & 0.68 &  13.78   \\
158 & 2.0-2.5 & 0.15 & 4.79(0.07)(0.13) & 6.18(0.08)(0.41) & 6.05(0.09)(0.23) & 4.32(0.09)(0.43) & 0.57(0.01)(0.09) & 0.70 &  15.20   \\
158 & 2.0-2.5 & 0.24 & 4.46(0.14)(0.17) & 5.78(0.16)(0.60) & 4.60(0.14)(0.24) & 3.78(0.14)(0.33) & 0.54(0.03)(0.06) & 0.69 &  16.60   \\
158 & 2.0-2.5 & 0.34 & 4.65(0.40)(1.26) & 5.10(0.34)(1.34) & 3.16(0.26)(0.73) & 2.38(0.30)(0.20) & 0.41(0.07)(0.19) & 0.73 &  17.78   \\
\cline{2-10}
\hline

\noalign{\smallskip}\hline
\end{tabular}
}

\end{table*}

\section{Discussion}
\label{chap:discussion}
\subsection{Relation between radii and source parameters}

The correlation function of identical pions measured in heavy-ion collisions is determined by the space-time structure of the system at kinetic freeze-out. 
The conditions of the kinetic freeze-out result from the expansion of the dense and strongly interacting matter created in the collision 
into the surrounding vacuum.
The evolution of this expansion is defined by the properties of the matter itself;
in this way the equation of state of nuclear matter is linked to the measured correlations. 
  
The relation between the dynamic evolution of the system and the measured radii, characterizing the correlation function,
cannot be given in a model-independent way. Nevertheless, making several general assumptions, a set of equations connecting key parameters of
the expansion to the magnitude of the measured radii and their dependence on the transverse mass 
have been derived (see e.g. \cite{Wiedemann:1999qn}
and references therein).
With the transverse mass defined as $m_{t}=(m_{\pi}^2+k_{t}^2)^{1/2}$, where $m_{\pi}$ indicates the pion mass,
and $\beta_t = k_t / m_t$, the approximate relations are: 

\begin{eqnarray}
\label{equation:standardHBTrelations1}
&R_{long}& =  \tau_0 \cdot (T/m_t)^{1/2}  \\
\label{equation:standardHBTrelations2}
&R_{side}& =  R_{geo}/(1+m_{t}\cdot\eta_{f}^{2}/T)^{1/2} \\ 
\label{equation:standardHBTrelations3}
&R_{out}^{2}-R_{side}^{2}& =  \Delta\tau^{2} \cdot \beta_{t}^{2}.  
\end{eqnarray}

Here $\tau_0$ denotes the total lifetime of the system and $T$ the temperature, $R_{geo}$ the transverse
size of the pion source and $\eta_{f}$ the strength of the transverse expansion at freeze-out.
$\Delta\tau$ represents the duration of particle emission.
Although, Eqs.~\ref{equation:standardHBTrelations1}~-~\ref{equation:standardHBTrelations3} 
only hold under a number of assumptions, they may serve to guide the
interpretation of the measured correlation parameters.

The $R_{long}$ parameter is connected to the total lifetime of the system. 
In Figs.~\ref{fig:ktdep_5energies_midrap}~-~\ref{fig:rapdep_5energies2}
the $k_{t}$-dependence of $R_{long}$ changes only slightly with beam energy and rapidity interval.
Assuming a fixed freeze-out temperature this observation means that the lifetime of the system is independent of
pion pair rapidity and of incident beam energy.

Also the $k_{t}$-dependence of $R_{side}$ and the intercept when extrapolated to vanishing transverse momentum 
change little with beam energy, consequently the transverse radius of the particle source at freeze-out
 and the strength of the transverse expansion vary only weakly with beam energy. Closer inspection of
Fig.~\ref{fig:rapdep_5energies1} shows a systematic decrease of the $R_{side}$ parameter at low
transverse momenta with increasing rapidity, suggesting a smaller transverse radius at larger rapidities.

At lower $k_{t}$, the parameter $R_{out}$ is always larger than $R_{side}$, indicating a finite emission duration. 
In some cases at large transverse momenta, the radius parameter $R_{out}$ is smaller than $R_{side}$, rendering relation
~\ref{equation:standardHBTrelations3} meaningless. At RHIC energies some analyses also found
smaller values of $R_{out}$ compared to $R_{side}$ \cite{Adler:2001zd}, triggering several detailed 
theoretical studies,  e.g. \cite{Molnar:2002ax}.

Relations~\ref{equation:standardHBTrelations1}~-~\ref{equation:standardHBTrelations3} suffer from the fact, 
that only two-particle correlation functions are considered, while other observables 
related to the collective behaviour of the fireball are disregarded. 
Single particle spectra and anisotropic flow measurements also carry information
on the expansion and freeze-out temperature and should therefore be addressed together with the BE correlations
in order to get a consistent and complete picture of the evolution of the particle-emitting source.
In fact, the ambiguity of transverse expansion and temperature in Eqs.~\ref{equation:standardHBTrelations1} 
and~\ref{equation:standardHBTrelations2} can only be overcome if additional information is provided, 
e.g. by a simultaneous fit of single particle spectra and
the $k_{t}$-dependence of the correlation radii, as performed for the
first time in \cite{Appelshauser:1997rr}. 
Below, the radii will be compared to more comprehensive model calculations, which indeed
aim to describe not only the two-particle correlations but also other aspects of heavy-ion collisions.

\subsection{Energy dependence of BE parameters} 

\begin{figure*}
\centering
\resizebox{0.99\textwidth}{!}{  
\includegraphics{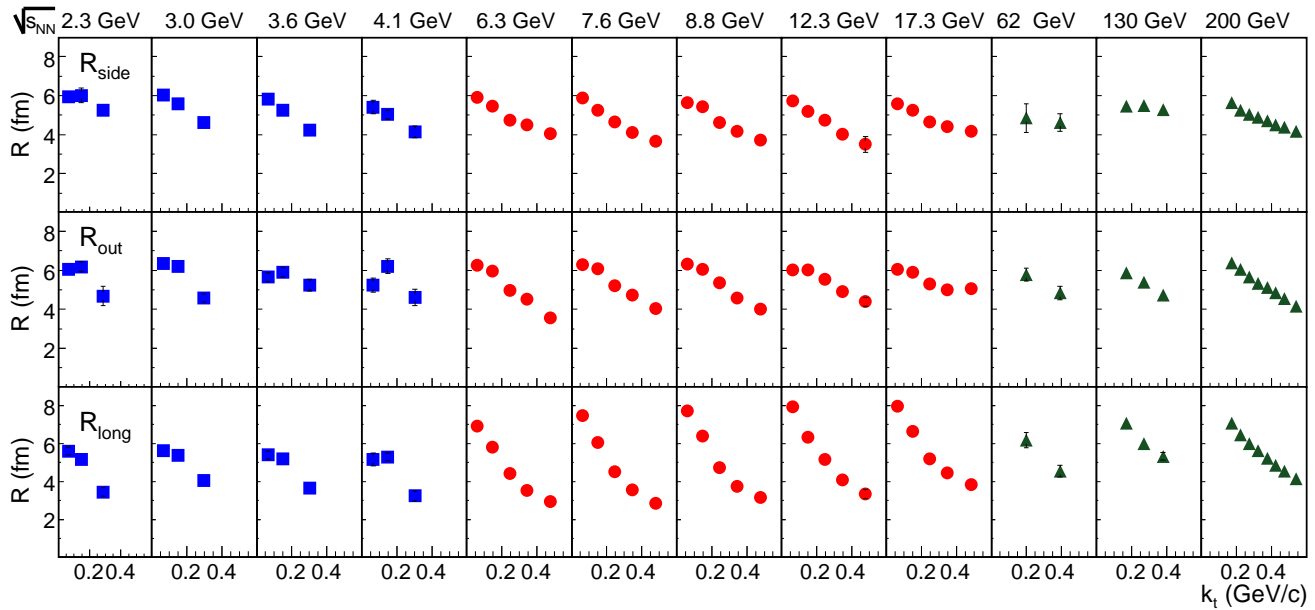}}
\caption{(Color online)~Summary of the $k_{t}$ dependence of the radii
  at various center of mass energies at midrapidity. Data are from AGS \cite{Lisa:2000hw} (squares), 
        this NA49 analysis (circles), PHOBOS \cite{Back:2004ug} (triangles at $\sqrt{s_{NN}}=62$~GeV), 
        and STAR \cite{Adler:2001zd,Adams:2004yc} (triangles at $\sqrt{s_{NN}}=130,200$~GeV).}
\label{fig:world_data}
\end{figure*}

The dependence of the parameters $R_{side}$, $R_{out}$, and $R_{long}$ 
on transverse momentum in central heavy-ion collisions at various center of mass energies is 
summarized in Fig.~\ref{fig:world_data}.
All the data correspond to measured pion-pion correlations at midrapidity, though it should be noted that the analyses
differ in details such as the choice of the transverse momentum variable and the treatment of the
Coulomb interaction.

\begin{figure}[htb]
\centering
\resizebox{0.45\textwidth}{!}{  
\includegraphics{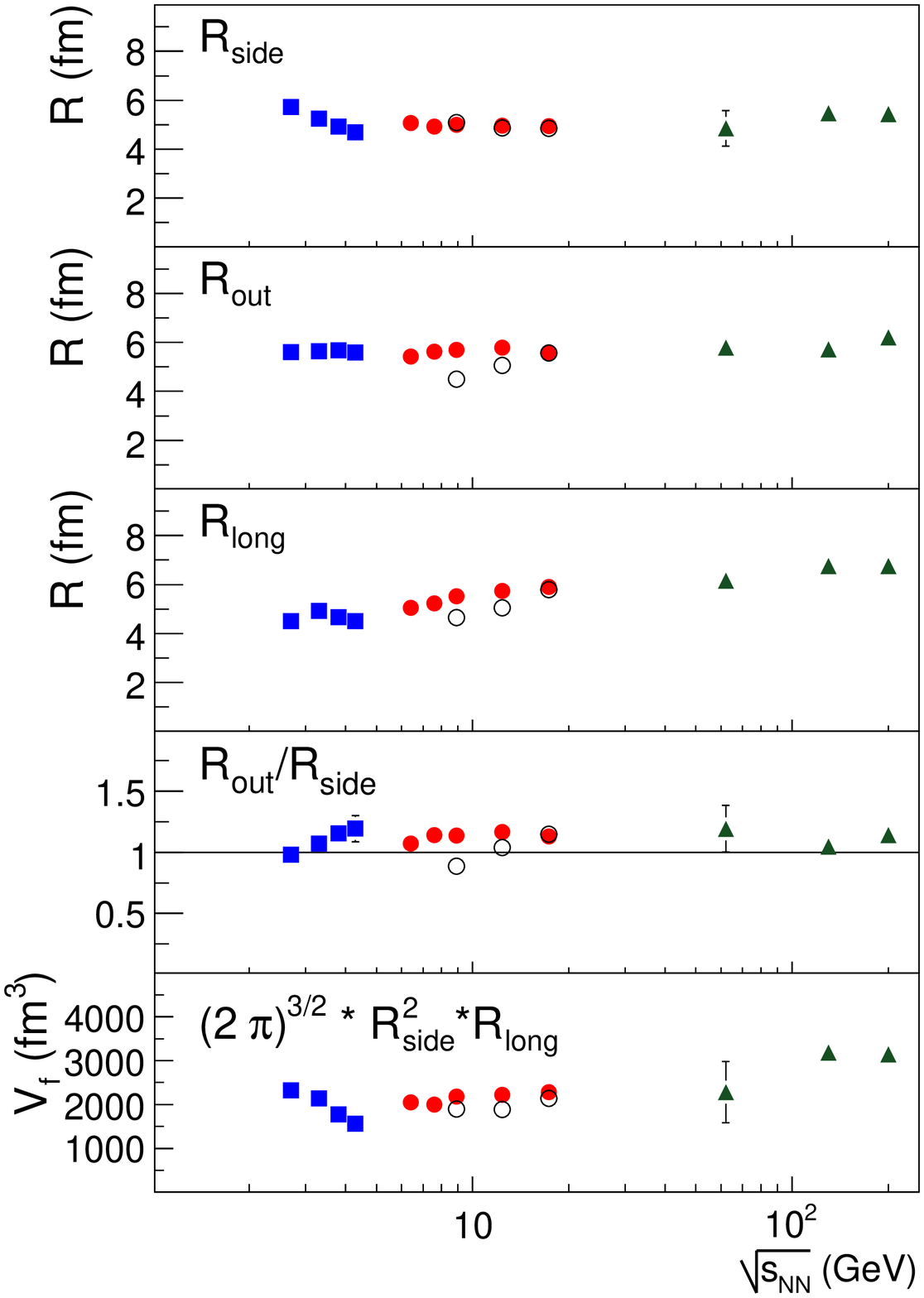}}

\caption{(Color online)~Dependence of the BE radii, the ratio  $R_{out}/R_{side}$ and the volume $V_{f}$ in central PbPb (AuAu) collisions 
        on beam energy at midrapidity. The $k_{t}$-dependence (Fig.~\ref{fig:world_data}) has been interpolated to get numerical values 
        of the radii at $k_{t}=0.2$ GeV/c.
\label{fig:world_data_at_200MeV}}
\end{figure}

From Fig.~\ref{fig:world_data} it is obvious that both the absolute values and the rate of decrease of the 
radii with increasing transverse momentum exhibit only a weak energy dependence. Since
correlations are assumed to reflect the evolution of the system and its structure at freeze-out,
these system properties therefore seem to depend only weakly on the initial conditions of the system.
As only central collisions are considered, the initial overlap region is similar, but 
the energy density increases by an order of magnitude between AGS and RHIC energies. 
This increase of the initial energy density causes a pronounced change 
in other observables, e.g. a strong increase of the particle abundances and a change in
the particle composition.

Fig.~\ref{fig:world_data_at_200MeV} shows the energy dependence of the BE radii at $k_{t}=0.2$~GeV/c. 
The plotted values have been calculated by linear interpolation of the data in Fig.~\ref{fig:world_data}.
The parameter $R_{side}$ decreases at AGS energies and remains almost constant
at higher energies, the $R_{out}$ parameter remains almost constant, only the $R_{long}$ parameter shows a trend to 
increase with the center of mass energy.

In many model calculations the parameter $R_{side}$ at $k_{t}=0$ is found to be closely related to the transverse size 
of the system. Under this assumption, the apparent independence of $R_{side}$ on the center of mass energy 
in Fig.~\ref{fig:world_data_at_200MeV} 
means that the system at freeze-out reaches roughly the same radial extension irrespective of the initial energy density. 

Provided Eq.~\ref{equation:standardHBTrelations1} is a valid approximation and the freeze-out temperature is constant,
the slight increase of the $R_{long}$ parameter might be connected to a longer overall lifetime of the
pion source created in central collisions at larger center of mass energies. 

The ratio $R_{out}$ to $R_{side}$ has been related to the emission duration in the emission function. According 
to Fig.~\ref{fig:world_data_at_200MeV} this ratio is close to unity independent of the center of mass energy.
This corresponds (see Eq.~\ref{equation:standardHBTrelations3}) to a small emission duration, i.e. to a sudden freeze-out.

In \cite{Adamova:2002ff} it has been suggested that the freeze-out occurs when the mean free path
of the pions, 
\begin{equation}
\lambda_{f} = {{V_{f}} \over { N \cdot \sigma }}, 
\end{equation}
reaches a value of about 1 fm. Here, $N$ denotes the number of particles in the freeze-out volume
$V_{f}$ and $\sigma$ the pion cross section. 
The freeze-out volume has been related to the radii by
\begin{equation}
V_{f} = {{(2\pi)}^{3/2}} \cdot R_{side}^2 \cdot R_{long}.
\end{equation}
Fig.~\ref{fig:world_data_at_200MeV} shows this quantity in the lowest panel. In \cite{Adamova:2002ff}
a minimum in the energy dependence of $V_{f}$ was observed at lower SPS energies, where the system transforms
from a baryon-  to a meson-dominated system. The radii determined in the present analysis
do not confirm this result.  
 
Another ansatz for the interpretation of the beam energy dependence of the extracted radii is given in 
\cite{Akkelin:2004he,Akkelin:2005ms}. The product of the radii $R_{side}$, $R_{out}$, and $R_{long}$
is used to derive the phase space density of the system averaged over momentum and configuration space. 
Since this quantity is believed to be conserved at the later stages of the assumed isentropic
evolution of the system, it is closely related to the conditions of the system at earlier times, 
when possibly a phase transition from deconfined matter to a hadronic system occurred. 
It was found \cite{Akkelin:2005ms} that the averaged phase space density shows an energy dependence
consistent with the appearance of a deconfined phase at lower SPS energies.

\subsection{Rapidity dependence at SPS}

\begin{figure}[!htb]
\centering
\resizebox{0.45\textwidth}{!}{  
\includegraphics{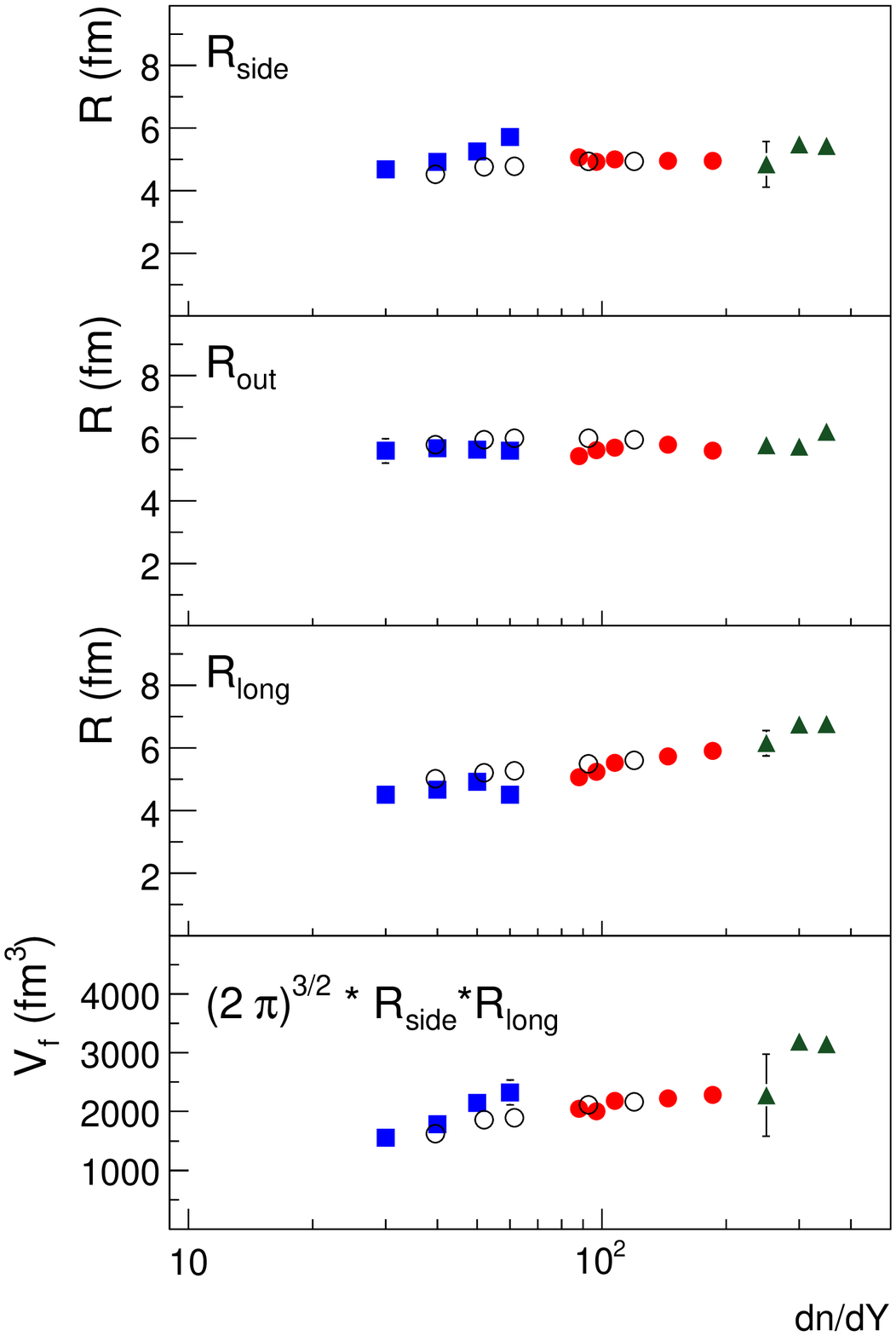}}
\caption{(Color online)~Dependence of the BE radii on $dn/dY$ in central PbPb (AuAu) collisions. 
        Closed symbol correspond to measurements at midrapidity,
        squares correspond to AGS energies, circles to SPS energies, and triangles to RHIC energies. 
        Open symbols represent NA49 data at forward rapidity $1.0<Y_{\pi\pi}<1.5$.
        All radii have been interpolated to get values at $k_{t}= 0.2$~GeV/c.}
\label{fig:hbt_vs_dndy}
\end{figure} 

This section discusses possible implications of the rapidity dependence of the BE correlation function
of pion pairs.  With the choice of a particular rapidity interval, particle pairs are selected that 
are emitted from certain parts of
the source: Assuming an emission function as described in Eq.~\ref{eq:Emissionfunction},
high pair rapidities correspond to large z-values of the last interaction (i.e. emission) point.
Using relations~\ref{equation:standardHBTrelations1}~-~\ref{equation:standardHBTrelations3}, 
the small decrease with increasing rapidity of the $R_{side}$ parameter at low transverse momentum 
implies a smaller transverse extension of the source as well as a higher expansion strength for the 
forward region of the system. Since $R_{out}$ does not change with increasing rapidity, but $R_{side}$ 
decreases slightly, the emission duration parameter, proportional to the difference of the two, 
also increases slightly. The $R_{long}$ parameter varies with rapidity, especially at low $k_{t}$, 
but not in a systematic way. Hence conclusions are hard to derive without consulting more refined 
model calculations.  The cross term $R_{outlong}$ clearly rises with increasing rapidity, 
indicating the strong longitudinal expansion. A different reference frame and parametrization 
as applied in \cite{Appelshauser:1997rr} is better suited to quantify the 
strength of the longitudinal expansion. The results in fact demonstrated approximate boost-invariance.

Experimentally a clear increase of the radii with decreasing impact parameters of the collision has been observed 
\cite{Adams:2004yc,Adamova:2002wi}.
Decreasing the impact parameter leads to larger rapidity densities. Therefore it has been speculated 
that there is
an explicit relation between the rapidity density and the size of the radii.
Since the rapidity density also depends on rapidity (see Fig.~\ref{fig:rap_pi_ka}), this quantity can also be varied 
-- at constant beam energy -- by changing the rapidity interval. Fig.~\ref{fig:hbt_vs_dndy} displays the dependence of the radii on the 
rapidity density in central heavy-ion collisions. Though the parameter $R_{long}$ increases with
increasing rapidity density, both $R_{side}$ and $R_{out}$ depend only little on rapidity density.
This contrasts with
the impact parameter dependence, where all three parameters increase from peripheral to central collisions.

\subsection{Blast-wave model}
\label{section:Lisa&Retiere}
In this section an analytical parametrization (blast-wave (BW) model) by Lisa and Retiere \cite{Retiere:2003kf} 
of the hadronic freeze-out is used to determine key characteristics of the 
evolution of the collision from the measured radii and transverse mass spectra.
The parametrization is based on a hydrodynamical ansatz, in which the expansion of the particle source is quantified
in the most simple case by only two parameters: The transverse expansion velocity and the temperature.
This approach was introduced by Schnedermann et al. \cite{Schnedermann:1993ws} 
and has been further developed by several groups, e.g. \cite{Csorgo:1995bi,Wiedemann:1998ta,Tomasik:2001uz,Huovinen:2001cy}.
Details of the parametrization applied here are given in \cite{Retiere:2003kf}.

The model is based on the assumption of a classical emission function $S(x,p)$,
\begin{eqnarray}
\label{eq:Emissionfunction}
S(x,p) &=&  m_t\cosh(\eta - y) \exp \left( {\frac{-(\tau-\tau_0)^2}{2\Delta\tau^2}} \right ) \nonumber  \\
&&\cdot \Omega(r) \left( \exp \left( {{-p\cdot u(x)/T}+s} \right ) \right ) ^{-1},
\end{eqnarray}
which represents the probability that a pion of momentum $p$ decouples at the space point $x$ 
from the particle source. The emission function in form of Eq.~\ref{eq:Emissionfunction} is valid only in the 
longitudinal co-moving system. 
On the right hand side of Eq.~\ref{eq:Emissionfunction} the transverse mass of the emitted particle 
is labeled $m_{t}$, the space-time rapidity of the emission point $\eta = \frac{1}{2}\ln\left[(t+z)/(t-z)\right]$,
and the particle rapidity $y$. If the particle obeys Bose-Einstein statistics $s$ is
set to $s=-1$, if Fermi-Dirac statistics has to be applied it is set to $s=+1$.  

The emission is assumed to occur according to a Gaussian distribution in longitudinal proper 
time $\tau = \sqrt{t^{2}-z^{2}}$ with the mean value labeled $\tau_{0}$ and the standard deviation $\Delta\tau$.
The density distribution of the source is modeled as a uniform cylinder, i.e. $\Omega({\sqrt{x^2+y^2}}) = \Omega(r) $ 
is set to unity if $r<R_{geo}$  and zero otherwise (so called 'box profile').  

The Boltzmann factor $\exp(-p \cdot u(x)/T)$ arises from the assumption of local thermal equilibrium 
at a temperature $T$ within a source element moving with four velocity $u(x)$.
In the transverse plane, the flow rapidity is assumed to increase linearly with the distance from the origin
(beam axis), and the maximum transverse rapidity $\rho_0$ at the cylinder surface $R_{geo}$
is treated as a free parameter. 
The longitudinal flow velocity is set to $v_L = z/t$ as required by longitudinal boost-invariance.
This assumption, introduced by Bjorken \cite{Bjorken:1982qr}, simplifies the model, 
but the finite extension of the system and possible deviations from a boost-invariant expansion are neglected.

Fixing the values of the free parameters in Eq.~\ref{eq:Emissionfunction} and integrating 
it yields the predictions for the single-particle spectra:

\begin{equation}\frac{dN}{p_t dp_t}  \propto  \int d^4x S(x,p).
\label{eq:spectra}
\end{equation}

\noindent An optimal set of parameters can be obtained by comparing these predictions
with measured transverse momentum spectra and 
employing a fit procedure. Here it should be pointed out, that the BW parametrization neglects the fact
that a large fraction of the measured hadrons, particularly low-$p_t$ pions, stems from resonance decays. 
The measured spectra actually represent the sum of the spectra of the primary particles and the 
decay spectra of strongly decaying resonances.
Experimentally, these two contributions  can not be distinguished.
Effects of resonance feeddown on fits with the BW model were minimized by using pion spectra 
only for $m_t > 0.4$~GeV/c.

In the BW model \cite{Retiere:2003kf} the radius parameters are obtained by calculating numerically
the appropriate spatial and temporal moments of the emission function $S(x,p)$
for a given set of BW parameter values in Eq.~\ref{eq:Emissionfunction}.
Using an iterative fit procedure, 
the optimal parameters of the emission function can then be determined by comparison
of the computed radius parameters with their measured values and $k_{t}$-dependence.

The fitting procedure can treat single-particle spectra and radius parameters simultaneously, ensuring that the
finally extracted emission function is better constrained than in studies which consider only one
set of observables.
 
In the present application of the model an azimuthally symmetric source is assumed, since only central collisions are studied. 
With this restriction the model features five free parameters: the freeze-out temperature $T$, the maximum transverse rapidity $\rho_{0}$,
the freeze-out transverse radius $R_{geo}$, the overall lifetime of the source $\tau_0$ and the emission duration 
$\Delta \tau$.             
Elliptic flow measurements are not included in this study, since the analysis focuses on central
collisions. In the following, this parametrization will be fitted to radius parameters and to
transverse mass spectra at midrapidity using only statistical errors.

\begin{figure}[!th]
\centering
\resizebox{0.45\textwidth}{!}{
\includegraphics{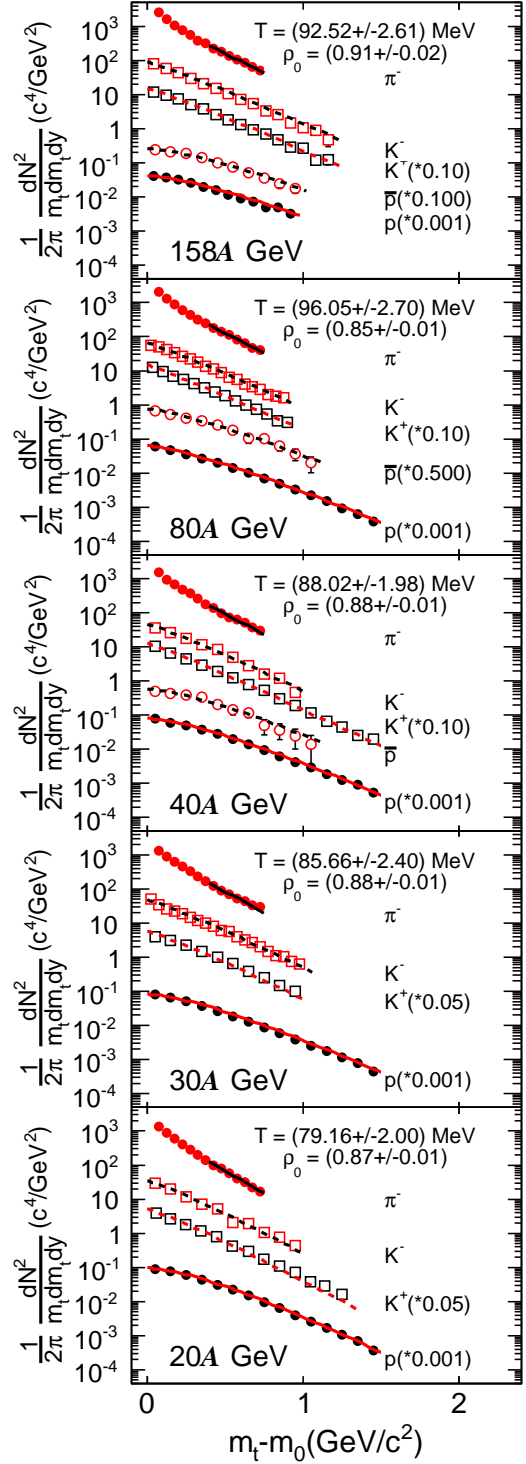}}

\caption{(Color online)~Transverse mass spectra of pions, kaons, and protons in central PbPb collisions at 
midrapidity measured by NA49. m$_0$ is the particle rest mass.
Lines indicate the fit results, numerical values of the fit parameters $T$ and $\rho_{0}$ are also shown.
Further details are given in the text.}
\label{fig:spec_20_30}
\end{figure}

The data points in Fig.~\ref{fig:spec_20_30} show single-particle spectra measured by NA49 
at five beam energies at midrapidity and the lines indicate the fit results. 
The $m_{t}$ dependence of the production of the various particle species 
is well described by the parametrization. Since the model only aims to reproduce the 
shape of the spectra, the normalisation, i.e. the particle yield, was treated as a free parameter
in the fitting procedure.
Solid lines in Fig.~\ref{fig:spec_20_30} are obtained from the fitting
procedure when negative pions, protons and the radii are used as
input. The predictions for K$^+$, K$^-$ and antiproton spectra were then
calculated using the fitted parameters and are shown by the dashed
curves in Fig.~\ref{fig:spec_20_30}. They are also in excellent agreement with the measurements.

\begin{figure}[!h]
\centering
\resizebox{0.45\textwidth}{!}{
\includegraphics{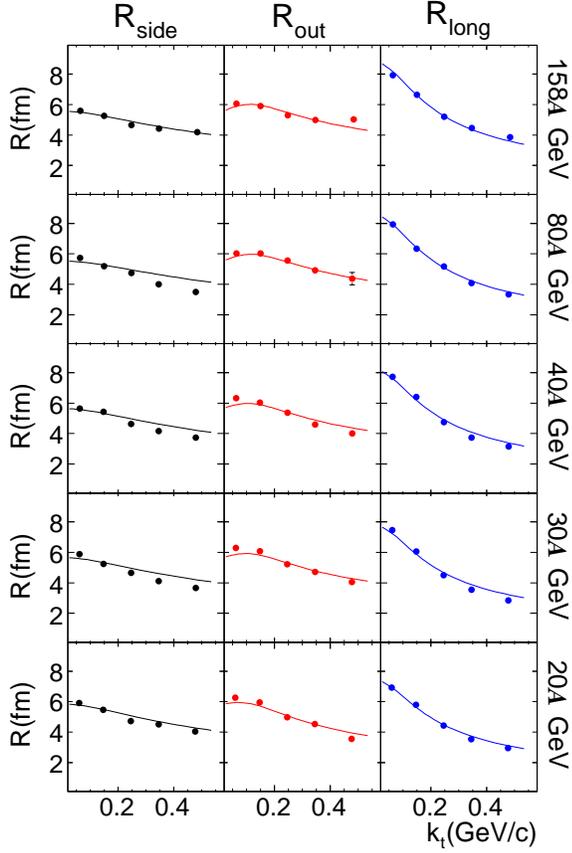}}
\caption{\label{fig:par_same_spec} (Color online)~$k_{t}$-dependence of the radius parameters
 in central PbPb collisions 
        at midrapidity (data points) compared to fit results (lines). }
\end{figure}

Fig.~\ref{fig:par_same_spec} shows the $k_{t}$ dependence of the
parameters $R_{side}$, $R_{out}$, and $R_{long}$ 
at different incident beam energies at midrapidity compared with the fit values. Both, the absolute values and the transverse
momentum dependences are well reproduced. Only the gradient of the $k_t$-dependence of the parameter $R_{side}$ tends to be smaller
in the parametrization than in the measurement.

\begin{figure}[!b]
\centering
\resizebox{0.45\textwidth}{!}{   
\includegraphics{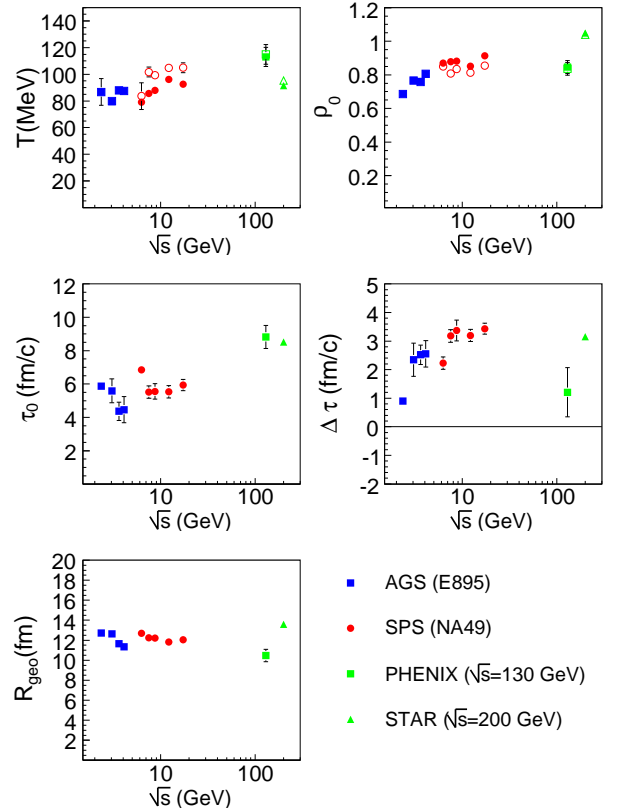}}
\caption{(Color online)~Energy dependence of the model parameters $T$, $\rho_{0}$, 
        $\tau_0$, $\Delta\tau$, and $R_{geo}$.
        Input to the fit procedure are the spectra of $\pi^{-}$ and  protons, and the radii.
        Open symbols correspond to a fit of only the spectra, disregarding the radii. 
        Further details and references are given in the text.
}
\label{fig:endep_sourcepar_same_spec}
\end{figure}

The energy dependence of the extracted BW model parameters is plotted in Fig.~\ref{fig:endep_sourcepar_same_spec}.
In addition to results obtained from fits to NA49 data at SPS energies, the figure also includes parameters 
derived by the same fit procedure from data at lower energies at the AGS \cite{Klay:2002,Klay:2003,Lisa:2000hw}
and at higher energies at RHIC \cite{Adams:2004yc,Adams:2004pip,Adcox:2002hbt,Adcox:2002pip}. 
From experiments at the AGS radii for pions and transverse momentum spectra of only pions 
and protons are available. For consistency, also the fits at higher energies used as input only radii and 
transverse momentum spectra of pions and protons.
Open symbols in Fig.~\ref{fig:endep_sourcepar_same_spec} represent the result when
only pion and proton spectra are fitted
and two-particle correlations are disregarded. In this special case only two parameters are determined, $T$ and
$\rho_{0}$. 

The temperature parameter tends to increase with the available center of mass energy, whereas the 
surface expansion rapidity $\rho_{0}$ saturates at lower SPS energies.
The kinetic freeze-out temperature of about 80-100 MeV is lower than the hadro-chemical freeze-out temperature 
at an earlier stage of the collision. The latter can be derived from the particle species composition
resulting in values of about 140-160 MeV \cite{Becattini:2003wp}.
This observation complies with the assumption that the system cools down by rapid expansion first
through chemical freeze-out (when inelastic collisions stop in the fireball) and then through
kinetic freeze-out (when also elastic interactions cease).

The transverse radius at freeze-out is independent of the beam energy at about 12 fm, which is approximately twice the size of a lead nucleus,
indicating the growth of the system.
The lifetime of the source depends weakly on energy; it is about 4-6 fm/c at lower energies and increases to 8 fm/c at RHIC energies.

The emission duration takes a finite value at lower energies and
increases up to the top SPS energy, then it saturates at a value of $\approx$~3 fm/c.

\subsection{Comments on other models}
\label{section:othermodels}
The Buda-Lund collaboration uses a parametrization similar to the BW model described in the previous section
to describe simultaneously single-particle spectra and radii. The parametrization is introduced in detail in 
\cite{Csorgo:1995bi,Ster:1998hu} and has been successfully applied to SPS and 
RHIC data \cite{Ster:1999ib,Csanad:2003sz}.
The main extensions compared to the simpler blast-wave model 
are the assumption of a local temperature $T(x)$ instead of a single global temperature $T$, 
and the introduction of a local chemical potential $\mu(x)$.
Preliminary results using this parametrization are similar to those obtained with the BW model discussed
in the previous section. A detailed analysis of the data presented here is in progress.












A different approach is provided by pure hydrodynamic calculations, an idealized but well-defined limiting case scenario
for the evolution of the system generated in central heavy-ion collisions.
This kind of model relies only on the assumption of an equation of state, a set of initial conditions, and on a specific freeze-out criterion.
The matter is treated as an ideal, locally thermalized fluid whose dynamics are governed
by the relativistic hydrodynamic equations. 
The freeze-out is usually modeled to occur at a fixed temperature. 
If the local temperature
of a fluid cell drops below the freeze-out temperature, the properties of this cell are converted into
an ideal gas of hadrons and hadronic resonances (Cooper-Frye prescription) \cite{Tomasik:2002rx,Cooper:1974mv}.
This leads to an emission function which can be used to calculate the two-particle correlations.
This class of models successfully describes bulk properties like the elliptic flow and the
transverse momentum spectra of pions. However, the models often fail to reproduce the
experimentally observed $k_{t}$-dependence of the radii \cite{Zschiesche:2001dx,Morita:2002av}.
The overall lifetime in this class of models is usually found to be large, hence the 
$R_{long}$ parameter is often overpredicted.
Depending on the assumed equation of state and initial conditions,
the $R_{side}$ parameter is often underpredicted and $R_{out}$ is overpredicted.
Consequently, the ratio of these two parameters, related to the emission duration,
is not well reproduced.
If the system passes through a first-order phase transition the $R_{out}$ parameter was predicted to 
become much larger than $R_{side}$ \cite{Rischke:1996em}.
Experimentally this signature has not been observed, 
as can be seen from Fig.~\ref{fig:world_data_at_200MeV}. 
However, more refined models are able to describe the data. E.g., in \cite{Borysova:2005ng} a protracted
surface emission is introduced instead of the conventional Cooper-Frye prescription \cite{Cooper:1974mv}. In this
case the $k_{t}$-dependence of the ratio $R_{out} / R_{side}$ is well reproduced.
The failure of pure hydrodynamic models might also be due to the fact that the correlations are determined at the 
final stages of the evolution, when the mean free path of the
particles is finite, possibly of the order of the size of the system. 
In contrast, the basic assumption of the hydrodynamic approach is a zero mean free path.




A different concept is pursued by microscopic models. 
These provide an approach complementary to the hydrodynamics inspired ansatz described above.
In transport approaches, the heavy-ion collisions are modeled by initial particle production
from string fragmentation followed by evolving these particles along straight-line trajectories punctuated by
collisions according to free-space cross-sections. The result of this microscopic simulation is the state of
the system at freeze-out, i.e. the momenta and space-time points of the last interaction. 
A well known implementation of this kind of models is UrQMD,  which is 
described in detail in \cite{Bass:1998ca,Bleicher:1999xi,Bratkovskaya:2004kv}.
Quantum statistics particle correlations are in principle not included in transport models. 
Therefore, BE correlations have to be introduced in a second step, e.g.
by using an afterburner like the CRAB algorithm \cite{Pratt:1994uf,Pratt:2006aa}. 

The UrQMD model combined with the CRAB algorithm was compared recently to the radius
parameters measured at SPS energies \cite{Li:2006gb}.
The model is able to reproduce qualitatively the $k_{t}$ and rapidity dependence of the radii 
at all energies. In particlular it reproduces the weak rapidity dependence of the parameters 
$R_{side}$, $R_{out}$, and $R_{long}$, and the strong increase of the cross term
$R_{outlong}$ with increasing forward rapidity.
The parameter $R_{side}$ is slightly underestimated and $R_{out}$ is overestimated at
larger transverse momenta. Consequently, the ratio $R_{out} / R_{side}$ exceeds
the measured values.

\section{Conclusion}
This paper presents a detailed study of $\pi^{-}\pi^{-}$ Bose-Einstein correlations in 
central heavy-ion collisions at five SPS energies. The $k_{t}$ and rapidity dependence 
of the fit parameters $\lambda$, $R_{side}$, $R_{out}$, $R_{long}$, and 
$R_{outlong}$ is obtained.
A decrease of the radius parameters with increasing transverse momentum is observed.
This behaviour has been reported before and is usually attributed to the rapid
expansion of the pion source.
 
Absolute values as well as the gradient of the transverse momentum dependence do not change 
with incident beam energy. A blast-wave parametrization of the evolution of the system 
motivated by hydrodynamics yields a freeze-out temperature of 80-100 MeV, a transverse expansion
velocity of the order of 72 \% of the speed of light at the surface, a transverse radius of the system 
of about 12 fm, a total lifetime of about 6 fm/c and a finite emission duration of about 3 fm/c.
The energy dependence of the radius parameters as well as of the fitted blast-wave model parameters 
shows no particular structure.

The study of the rapidity dependence of BE correlations reveals that
only the $R_{side}$ parameter depends significantly on the selected pair rapidity.
This parameter decreases systematically from midrapidity to forward rapidity. $R_{out}$ depends only
weakly on rapidity and $R_{long}$ shows some variations at low transverse momenta, but not 
in a systematic way. The weak rapidity dependence of the radii contrasts with the strong change of the 
rapidity density of produced particles.

The data set presented here provides stringent constraints for models trying to describe the
evolution of central heavy-ion collisions at SPS energies. Experiment NA61, an upgrade
and continuation of the NA49 experiment, will extend the systematics with measurements 
of smaller collision systems \cite{Antoniou:2006mh}. Moreover, an analysis of BE correlations
in NA49 using the source imaging technique \cite{Chung:2007si} is in progress.

\section{Acknowledgments}
This work was supported by the US Department of Energy
Grant DE-FG03-97ER41020/A000,
the Bundesministerium fur Bildung und Forschung, Germany (06f137), 
the Virtual Institute VI-146 of Helmholtz Gemeinschaft, Germany,
the Polish State Committee for Scientific Research (1 P03B 006 30, 1 P03B 097 29, 1 PO3B 121 29, 1 P03B 127 30),
the Hungarian Scientific Research Foundation (T032648, T032293, T043514),
the Hungarian National Science Foundation, OTKA, (F034707),
the Polish-German Foundation, the Korea Science \& Engineering Foundation (R01-2005-000-10334-0) and the Bulgarian National Science Fund (Ph-09/05).

%
\bibliographystyle{h-physrev4.bst}
\bibliography{paper}
%
%
%

\end{document}